\documentclass[format=acmsmall, review=false, screen=true]{acmart}

\usepackage{graphicx}
\usepackage{subcaption}
\usepackage{caption}
\usepackage{booktabs} 
\usepackage{fancyvrb}
\usepackage{pifont}
\usepackage{graphics}
\usepackage{multirow}

\usepackage[ruled]{algorithm2e} 

\newcommand{\Srndic}{\v{S}rndi\'{c}\xspace}

\newcommand{\ie}{\emph{i.e.}\xspace}
\newcommand{\eg}{\emph{e.g.}\xspace}
\newcommand{\etal}{\emph{et al.}\xspace}

\newcommand{\vct}[1]{\ensuremath{\boldsymbol{#1}}} 

\newcommand{\set}[1]{\ensuremath{\mathcal{#1}}}
\newcommand{\con}[1]{\ensuremath{\mathsf{#1}}}

\newcommand{\argmin}{\operatornamewithlimits{\arg\,\min}}

\SetAlFnt{\small}
\SetAlCapFnt{\small}
\SetAlCapNameFnt{\small}
\SetAlCapHSkip{0pt}
\IncMargin{-\parindent}
\newfloat{Listing}{htbp}{loa}
\newcommand{\myparagraph}[1]{\smallskip \noindent \textbf{#1.}}

\setcopyright{acmcopyright}
\acmJournal{CSUR}
\acmYear{2019} \acmVolume{1} \acmNumber{1} \acmArticle{1} \acmMonth{1} \acmPrice{15.00}\acmDOI{10.1145/3332184}

\begin{document}

\title{Towards Adversarial Malware Detection: Lessons Learned from PDF-based Attacks}

\author{Davide Maiorca}
\affiliation{%
  \institution{University of Cagliari}
  \streetaddress{Piazza d'Armi}
  \city{Cagliari}
  \postcode{09123}
  \country{Italy}}
\email{davide.maiorca@diee.unica.it}
\author{Battista Biggio}
\affiliation{%
  \institution{University of Cagliari}
\streetaddress{Piazza d'Armi}
\city{Cagliari}
\postcode{09123}
\country{Italy}}
\affiliation{%
    \institution{Pluribus One} 
    \country{Italy}}
\email{battista.biggio@diee.unica.it}
\author{Giorgio Giacinto}
\affiliation{%
	\institution{University of Cagliari}
	\streetaddress{Piazza d'Armi}
	\city{Cagliari}
	\postcode{09123}
	\country{Italy}}
\affiliation{%
    \institution{Pluribus One}
    \country{Italy}}
\email{giacinto@unica.it}

\begin{abstract}
Malware still constitutes a major threat in the cybersecurity landscape, also due to the widespread use of infection vectors such as documents. These infection vectors hide embedded malicious code to the victim users, facilitating the use of social engineering techniques to infect their machines.
Research showed that machine-learning algorithms provide effective detection mechanisms against such threats, but the existence of an arms race in adversarial settings has recently challenged such systems.
In this work, we focus on malware embedded in PDF files as a representative case of such an arms race.
We start by providing a comprehensive taxonomy of the different approaches used to generate PDF malware, and of the corresponding learning-based detection systems.
We then categorize threats specifically targeted against learning-based PDF malware detectors, using a well-established framework in the field of adversarial machine learning.
This framework allows us to categorize known vulnerabilities of learning-based PDF malware detectors and to identify novel attacks that may threaten such systems, along with the potential defense mechanisms that can mitigate the impact of such threats.
We conclude the paper by discussing how such findings highlight promising research directions towards tackling the more general challenge of designing robust malware detectors in adversarial settings. 
\end{abstract}



\maketitle

\renewcommand{\shortauthors}{D. Maiorca, B. Biggio, G. Giacinto}

\section{Introduction}

Malware for X86 (and more recently for mobile architectures) is still considered one of the top threats in computer security. While it is common to think that the most dangerous attacks are crafted using plain executable files (especially on Windows-based operating systems), security reports showed that the most dangerous attacks in the wild were carried out by using \emph{infection vectors}~\cite{symantec18-april-rep}. With this term, we define non-executable files whose aim is exploiting vulnerabilities of third-party applications to trigger download (or direct execution) of executable payloads. Using such vectors gives attackers multiple advantages. First, they can exploit the structure of third-party formats to conceal malicious code, making their detection significantly harder. Second, infection vectors can be effectively used in social engineering campaigns, as victims are more prone to receive and open documents or multimedia content. Finally, although vulnerabilities of third-party applications are often publicly disclosed, they are not promptly patched. The absence of proper security updates makes thus the lifespan of attacks perpetrated with infection vectors much longer. 

Machine learning-based technologies have been increasingly used both in academic and industrial environments (see \eg,~\cite{kaspersky17-rep}) to detect malware embedded in infection vectors like malicious PDF files. Research work has demonstrated that learning-based systems could be effective at detecting obfuscated attacks that are typically able to evade simple heuristics~\cite{maiorca12-mldm,smutz12-acsac,srndic13-ndss,corona14-aisec}, but the problem is still far from being solved. Despite the significant increment of detected attacks, researchers started questioning the reliability of learning algorithms against \emph{adversarial} attacks carefully-crafted against them~\cite{barreno06-asiaccs,bruckner12-jmlr,biggio13-ecml,biggio14-svm-chapter,biggio18-pr}. Such attacks became widely popular when researchers showed that it was possible to evade deep learning algorithms for computer vision with \emph{adversarial examples}, \ie, minimally-perturbed images that mislead classification~\cite{szegedy14-iclr,goodfellow15-iclr}.
The same attack principles have also been employed to craft \emph{adversarial malware samples}, as first shown in~\cite{biggio13-ecml}, and subsequently explored in~\cite{srndic14-sp,wang17-kdd,demontis17-tdsc,grosse17-esorics,kolosnjaji18-eusipco}. 
Such attacks can typically perform few, fine-grained changes on correctly detected, malicious samples to have them misclassified as legitimate. Accordingly, it becomes possible to evade machine-learning detection in a stealthier manner, without resorting to invasive changes like those performed through code obfuscation.

Malicious PDF files constitute the most studied infection vectors in adversarial environments~\cite{smutz12-acsac,maiorca13-asiaccs,biggio13-ecml,srndic14-sp,maiorca15-icissp,maiorca15-chapter,carmony16-ndss,xu16-ndss,smutz16-ndss,maiorca19-sp}. This file type was chosen for three main reasons. First, it represented the most dangerous infection vector in the wild from $2010$ to $2014$ (to be subsequently replaced by Office-based malware), and a lot of machine learning-based systems were developed to detect the vast variety of polymorphic attacks related to such format (\eg,~\cite{maiorca12-mldm,smutz12-acsac,smutz16-ndss}). Second, the complexity of the PDF file format allows attackers to employ various solutions to conceal code injection or other attack strategies. Finally, it is easy to modify its structure, by for example injecting benign or malicious material in various portions of the file. Such characteristic makes PDF files particularly prone to be used in adversarial environments, as the effort that attackers have to carry out to create attack samples is significantly low.

While previous surveys in the field of PDF malware analysis focused on describing the properties of detection systems ~\cite{elingiusti18-chapter,nissim15-cose}, our work explores the topic under the perspective of adversarial machine learning. The idea is showing how adversarial attacks have been carried out against PDF malware detectors by exploiting the vulnerabilities of their essential components, and highlighting how the arms race between attackers and defenders has evolved in this scenario over the last decade. The result is two-folded: on the one hand, we highlight the current security issues that allow attackers to deceive the current state-of-the-art algorithms; on the other hand, understanding adversarial attacks points out new, intriguing research directions which we believe can also be relevant for other malware detection problems.

To adequately describe PDF malware detection under adversarial environments, we organized our work as follows.  First, we describe the main attack types that can be carried out by using PDF files. Second, we provide a detailed description of state-of-the-art PDF detectors, by primarily focusing on their machine-learning components. 
Third, we show how such systems can be evaded with different adversarial attacks. In particular, we provide a complete taxonomy of the attacks that can be performed against learning-based detectors, by also describing how such attacks can be concretely implemented and deployed. 
Finally, we overview possible solutions that have been proposed to mitigate such attacks, thus sketching further research directions in developing novel adversarial attacks and defenses against them.  

This work also aims to provide solid bases to overcome the hurdles that can be encountered when working in adversarial environments. We firmly believe that these principles can constitute useful guidelines when dealing with the generic task of malware detection, not only restricted to PDF files. We claim that systems based on machine learning for malware detection should be built by accounting for the presence of attacks carefully tailored against them, \ie, according to the principle of \emph{security by design}, which suggests to proactively model and anticipate the attacker to build more secure systems~\cite{biggio14-tkde,biggio18-pr}. 

The rest of the paper is structured as follows: Section~\ref{sec:malware} provides a general overview of the attacks carried out with the PDF file format, as well as the possible attacks in the wild that can be carried out against PDF readers; Section~\ref{sec:detection} depicts the various detection methodologies that have been introduced in these years, by also discussing the achieved detection performances; Section~\ref{sec:adversarial} provides a complete taxonomy of the adversarial attacks that can be carried out by exploiting learning-based vulnerabilities, as well as an insight into how these attacks can be implemented; Section~\ref{sec:discussion} describes the current countermeasures adopted against adversarial threats, and sketches possible research directions for developing new attacks and defenses. Section~\ref{sec:conclusion} concludes the paper by also providing a comparison with previous works.  

\section{PDF Malware}
\label{sec:malware}
This Section aims to provide the basics to understand how PDF malware in the wild executes its malicious actions. To this end, we divided this Section into two parts. In the first part (Section \ref{sec:malware:overview}), we provide a comprehensive overview of the PDF file format. In the second part (Section \ref{sec:malware:exploit}), we describe how PDF malware uses the characteristics of its file format to exploit vulnerabilities. 

\subsection{Overview of PDF files}
\label{sec:malware:overview}

Digital documents, albeit different concerning the way their readers parse them, share some essential aspects. In particular, they can be typically represented as a combination of two components: a general \emph{structure} that outlines how the document contents are stored, and the \emph{file content} that describes the information that is properly visualized to the user (such as text, images, scripting code). 

\begin{figure}[ht]
	\centering
	\includegraphics[width=0.7\textwidth]{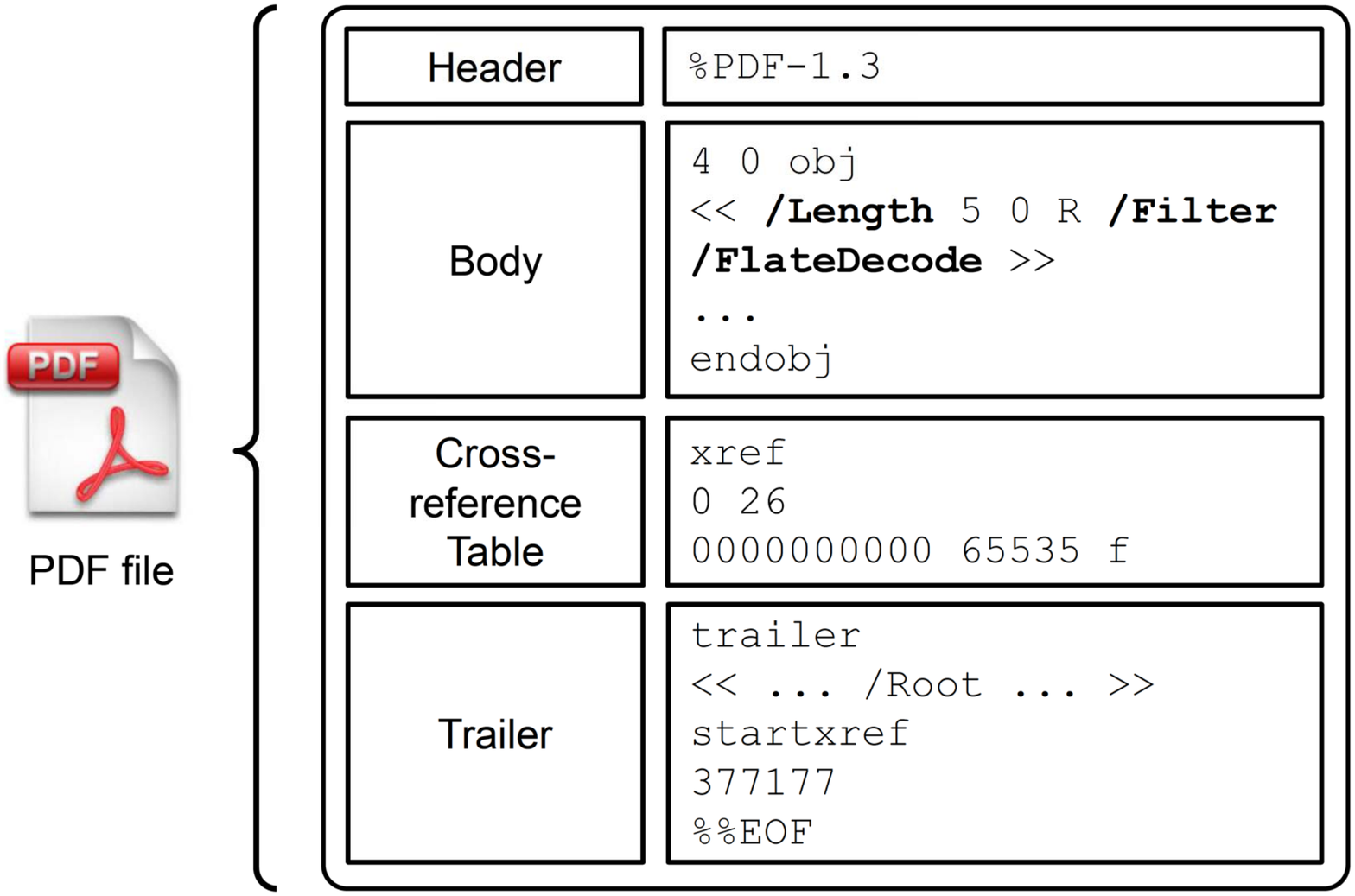}
	\caption{PDF file structure, with examples of header, body, cross-reference table and trailer contents. Object names (\ie, keywords) are highlighted in bold.}
	\label{sec:overview:fig:pdf-file}
\end{figure}

\myparagraph{General Structure} PDF (Portable Document Format) is one of the most used formats, along with Microsoft Office, to render digital documents. It can be conceptually considered as a graph of objects, each of them performing specific actions (\eg, displaying text, rendering images, executing code, and so forth). The typical structure of a PDF file is showed in Figure~\ref{sec:overview:fig:pdf-file}, and it consists of four parts~\cite{adobe06-reference,adobe08-supplement}: 

\begin{itemize}
	\item \textbf{Header}. A single line of text containing information about the PDF file version, introduced by the marker \%.
	\item \textbf{Body}. A sequence of objects that define the operations performed by the file. Such objects can also contain compressed or uncompressed embedded data (\eg, text, images, scripting code). Each object possesses a unique reference number, typically introduced by the sequence \texttt{number 0 obj}\footnote{The value between \texttt{number} and \texttt{obj} is called \emph{generation number}, and it is typically $0$. It can be different in some corner cases, but we recommend to refer to the official documentation for more information.}, where \texttt{number} is the proper object number. PDF objects can be referenced by others by using the sequence \texttt{number 0 R}\footnote{The value between \texttt{number} and \texttt{R} is the same generation number as the original object.}, where \texttt{number} identifies the \emph{target} object that is referenced. Each object is ended by the \texttt{endobj} marker. The functionality of each object is described by \emph{keywords} (also known as \emph{name objects} - highlighted in bold in the Figure), which are typically introduced by \texttt{/}.
	\item \textbf{Cross-Reference (X-Ref) Table}. A list of offsets that indicate the position of each object in the file. Such a list gives the PDF readers precise indications on where to begin parsing each object. The Cross-Reference Table is introduced by the marker \texttt{xref}, followed by a sequence of numbers, whose last indicates the total number of objects in the file. Each line of the table corresponds to a specific object, but only lines that end with \texttt{n} are related to objects that are concretely stored in the file. It is worth noting that all the PDF readers parse \emph{only the objects that are referenced by the X-Ref Table}. Therefore, it is possible to find objects that are stored in the file, but that lack their reference in the table. 
	\item \textbf{Trailer}. A special object that describes some basic elements of the file, such as the first object of the graph (\ie, where the PDF readers should start parsing the file information). Moreover, it contains references to the file metadata, which are typically stored in one single object. The keyword \texttt{trailer} always introduces the trailer object. 
\end{itemize}

PDF files are parsed as follows: first, the trailer object is parsed by PDF readers to retrieve the first object of the hierarchy. Then, each object of the PDF graph (contained in the body) is explored and rendered by combining information contained in the X-Ref Table with the references numbers found inside each object. Every PDF file is terminated by a \texttt{\%\%EOF} (End Of File) marker. An interesting characteristic of PDF files is that they can also be \emph{updated} without being regenerated from scratch (although the latter is possible), with a process called \emph{versioning}. When an existing object is modified, or new objects are added to the file, a new body, X-Ref table, and trailer are appended to the file. The new body and X-Ref table only contain information about the changes that occurred to the document.

\myparagraph{Objects} As previously stated, objects are typically identified by a number, and they are more formally referred to as \emph{indirect objects}. However, every element inside the body is generally regarded as an object, even if a number does not identify it. When an object is not identified by a number (\ie, when its part of other objects), it is called \emph{direct}. 

Indirect objects are typically composed of a combination of direct objects.  Listing \ref{sec:overview:subsec:objects:lis:obj} reports a typical example of PDF object.

\begin{center}
	\bigskip
	\begin{BVerbatim}
	8 0 obj
	<</Filter[/FlateDecode]/Length 384/Type/EmbeddedFile>>
	stream
	...
	endstream
	endobj
	
	\end{BVerbatim}
	\captionof{Listing}{An example of PDF object. The content of the stream has not been reported for the sake of brevity. }\label{sec:overview:subsec:objects:lis:obj}
	\bigskip
\end{center}

Most of the times, indirect objects (in this case, the object is number $8$) are dictionaries (enclosed in $<<$ $>>$) that contain \emph{sequences of coupled direct objects}. Each couple typically provides specific information about the object. The object in the Listing contains a sequence of three couples of objects, which decompress an embedded file (marked by the keywords \texttt{/Type/EmbeddedFile}) of length $384$ (keyword \texttt{/Length}) with a FlateDecode compression filter (keywords \texttt{/Filter/FlateDecode}). Objects that contain streams always feature the markers \texttt{stream} and \texttt{endstream} at the very end, meaning that PDF objects first instruct the PDF readers about their functionality, then on the data type they operate.  

Among the various types of objects, there are some that perform actions such as executing \texttt{JavaScript} code, opening embedded files, or even performing automatic actions when the file is opened or closed. To this end, the PDF language resorts to specific name objects (keywords) that are typically associated with actions that can have repercussions from a security perspective. Among the others, we mention \texttt{/JavaScript} and \texttt{/JS} for executing \texttt{JavaScript} code; \texttt{/Names}, \texttt{/OpenAction}, \texttt{/AA} for performing automatic actions; \texttt{/EmbeddedFile} and \texttt{/F} for opening embedded files; \texttt{/AcroForm} to execute forms. The presence of such objects should be considered as a first hint that somebody may perform malicious actions. However, as such objects are also widely used in benign files, it may be quite hard to establish the maliciousness of the file by only inspecting them.

\subsection{Malicious Exploitation} 
\label{sec:malware:exploit}

\begin{table*}[t]
	\centering
	\caption{An overview of the main exploited vulnerabilities against Adobe Reader, along with their vulnerability and exploitation type.} 
	\label{sec:attacks:tab:vulns}
	\begin{tabular}{@{}lccc@{}}
		\toprule
		\textbf{Vulnerability} & Vuln. Type & Exploitation Type \\ \midrule
		\textbf{CVE-2008-0655} & API Overflow (\emph{Collab.collectEmailInfo}) & JavaScript \\
		\textbf{CVE-2008-2992} & API Overflow (\emph{util.printf}) & JavaScript \\
		\textbf{CVE-2009-0658} & Overflow & File Embedding (JBIG2) \\
		\textbf{CVE-2009-0927} & API Overflow (\emph{Collab.getIcon}) & JavaScript \\
		\textbf{CVE-2009-1492} & API Overflow (\emph{getAnnots}) & JavaScript \\
		\textbf{CVE-2009-1862} & Flash (Memory Corruption) & ActionScript \\
		\textbf{CVE-2009-3459} & Malformed Data (FlateDecode Stream) & JavaScript \\
		\textbf{CVE-2009-3953} & Malformed Data (U3D) & JavaScript \\
		\textbf{CVE-2009-4324} & Use-After-free (\emph{media.newPlayer}) & JavaScript \\
		\textbf{CVE-2010-0188} & Overflow & File Embedding (TIFF) \\
		\textbf{CVE-2010-1240} & Launch Action & File Embedding (EXE) \\
		\textbf{CVE-2010-1297} & Flash (Memory Corruption) & ActionScript \\
		\textbf{CVE-2010-2883} & Overflow (\emph{coolType.dll}) & JavaScript \\
		\textbf{CVE-2010-2884} & Flash (Memory Corruption) & ActionScript \\
		\textbf{CVE-2010-3654} & Flash (Memory Corruption) & ActionScript \\
		\textbf{CVE-2011-0609} & Flash (Bytecode Verification) & ActionScript \\
		\textbf{CVE-2011-0611} & Flash (Memory Corruption) & ActionScript \\
		\textbf{CVE-2011-2462} & Malformed U3D Data & JavaScript \\
		\textbf{CVE-2011-4369} & Corrupted PRC Component & JavaScript \\
		\textbf{CVE-2012-0754} & Flash (Corrupted MP4 Loading) & ActionScript \\
		\textbf{CVE-2013-0640} & API Overflow & JavaScript \\
		\textbf{CVE-2013-2729} & Overflow & File Embedding (BMP) \\
		\textbf{CVE-2014-0496} & Use-After-Free (\emph{toolButton}) & JavaScript \\
		\textbf{CVE-2015-3203} & API Restriction Bypass & JavaScript\\
		\textbf{CVE-2016-4203} & Invalid Font & File Embedding (TFF)\\
		\textbf{CVE-2017-16379} & Type Confusion (\emph{IsAVIconBundleRec6}) & JavaScript \\
		\textbf{CVE-2018-4990} & Use-After-Free (ROP chains) & JavaScript \\
		\bottomrule
	\end{tabular}
\end{table*}

The majority of attacks that are carried out using documents resort to scripting codes to execute malicious code. Therefore, after having described the general structure of PDF files, we now provide an insight into the security issues related to the contents that can be embedded in the file. To do so, we provide a taxonomy of the various attacks that can be perpetrated by using PDF files. In particular, we focus on attacks targeting Adobe Reader, the most used PDF reader in the wild. The common idea behind all attacks is that they exploit specific vulnerabilities of the Adobe Reader components, and in particular of its plugins. Vulnerabilities can be characterized by multiple exploitation strategies, which also depend on the targeted Reader component. 

Table~\ref{sec:attacks:tab:vulns} reports a list of the major Adobe Reader vulnerabilities \emph{that have been exploited in the wild} (either with proofs-of-concept or proper malware) in the last decade, along with a brief description of their type and exploitation strategies. Notably, we did not include variants of the same vulnerability in the Table, and we only focused on the most representative ones. Such list has been obtained by analyzing exploit databases, media sources, security bulletins and our own file database retrieved from the \texttt{VirusTotal} service ~\cite{virustotal,exploit-db,malware-tracker,vuldb,fortinet-16,eset-18,sentinelone-18}. 
According to what we represented in Table \ref{sec:attacks:tab:vulns}, there are three primary ways to perform exploitation: 

\begin{itemize}
	\item \textbf{JavaScript-based}. These vulnerabilities are exploited by \emph{exclusively} employing \texttt{JavaScript} code, and it is the most common way to perform exploitation. The attack code can be scattered through multiple objects in the file, or it can be contained in one single object (especially in older exploits).
	\item \textbf{ActionScript-based}. These vulnerabilities exploit the capabilities of Adobe Reader of parsing Flash (\texttt{ActionScript}) files, due to the proper integration between Reader and Adobe Flash Player. \texttt{ActionScript} code can be used in combination to \texttt{JavaScript} to attain more effective exploitation. 
	\item \textbf{File Embedding}. This exploitation technique resorts to external file types, such as \texttt{.bmp}, \texttt{.tiff} and \texttt{.exe}. Typically, the exploitation is triggered when specific libraries of Adobe Reader attempt to parse such files. It is also possible to embed other PDF files: however, this is not considered as an exploitation technique, but more as a way to conceal other attacks (see the next sections for more details).
\end{itemize}

From this list, it can be observed that, although numerous vulnerabilities are still disclosed on Adobe Reader, only a few have been recently exploited in the wild. Such a tiny number of exploited vulnerabilities reflects the fact that PDF files are now less preferred as exploitation vectors by attackers. However, things can unexpectedly change when new, dangerous vulnerabilities are disclosed (such as \texttt{CVE-2018-4990}).
In the following, we provide a more detailed description of the previously mentioned exploitation techniques, by also providing concrete examples from existing vulnerabilities.

\myparagraph{JavaScript-Based Attacks} \texttt{JavaScript}-based attacks are the most used ones in PDF files due to the massive support that the file format provided to this scripting language. In particular, Adobe introduced specific API calls that are only supposed to be used in PDF files (their specifications are contained in the official Adobe references~\cite{adobe07-js-reference}), and that can be exploited to execute unauthorized code on a machine. 

According to the vulnerability types contained in Table \ref{sec:attacks:tab:vulns}, multiple vulnerabilities can be exploited through \texttt{JavaScript} code. Such vulnerabilities can be organized in multiple categories, which we describe in the following:   

\begin{itemize}
	\item \textit{API-based Overflow}. This vulnerability type typically exploits wrong argument parsing for specific API calls that belong to PDF parsing library, thus allowing attackers to perform buffer overflow or ROP-based attacks\footnote{Return Oriented Programming, an exploiting strategy that leverages return instructions to create shellcodes.} to complete the exploit. Typical examples of vulnerable APIs are \texttt{util.printf} and \texttt{Collab.getIcon}, which were among the first ones to be exploited when PDF attacks started to be massively used. 
	\item \textit{Use-After-Free}. This vulnerability type is based on accessing memory areas that have been previously freed (and not re-initialized). Normally, such behavior would make the program crash. However, it could allow arbitrary code execution in a vulnerable program.
	\item \textit{Malformed Data}. This vulnerability type is triggered by compressed malformed data that get decompressed at runtime. Such data is typically stored in streams. 
	\item \textit{Type Confusion}. This vulnerability type may affect functions that receive void pointer as parameters. Typically, the pointer type is inferred by checking some bytes of the pointed object. However, such bytes can be manipulated to make even an invalid pointer type to be recognized as valid, thus allowing arbitrary code execution. 
\end{itemize}

A large number of vulnerabilities and exploitation types allow attackers to carry out malicious actions that are often not easy to detect. The evolution of the employed exploitation techniques becomes particularly evident if we observe the differences between the first API overflows (\eg, \texttt{CVE-2008-0655}) and the most recent attack strategies. We expect that this trend will further evolve with more sophisticated techniques.    

\myparagraph{ActionScript-Based Attacks} As PDF files can visualize Flash content through Adobe Reader support to Adobe Flash technologies, one way to exploit Reader is triggering vulnerabilities of its Flash component by embedding malicious ShockWave Flash (SWF) files and \texttt{ActionScript} code. Normally, such code is used in combination with \texttt{JavaScript}: while executing \texttt{ActionScript} triggers the vulnerability itself, the rest of \texttt{JavaScript} code carries out and finalizes the exploitation. This exploitation technique was particularly popular in $2010$ and $2011$.

In a similar way to what we described for \texttt{JavaScript}, multiple vulnerabilities can be exploited by using \texttt{ActionScript}. In the following, we describe the prominent ones.

\begin{itemize}
	\item \textit{Memory Corruption}. This vulnerability occurs when specific pointer values in memory get corrupted in a way that they point to other memory areas, controlled by the attacker. It is the most used way to exploit Flash components.
	\item \textit{ByteCode Verification}. This vulnerability allows attackers to execute code in uninitialized areas of memory. 
	\item \textit{Corrupted File Loading}. This vulnerability is triggered when parsing specific, corrupted video files.
\end{itemize}

Generally speaking, vulnerabilities that affect the Flash components of Adobe Reader are more complex to exploit than others. This is because two types of scripting code must be executed, and exploitation is typically carried out in two stages (first - \texttt{ActionScript} execution, then - \texttt{JavaScript} code). For this reason, attackers typically prefer exploits that are simpler to be carried out. Moreover, Flash-based technologies will be dismissed in some years, giving attackers further reasons not to develop further exploits.

\myparagraph{File Embedding} This vulnerability exploits the presence of embedded, malformed content inside the PDF file. Typically, decoding such content leads to automatic memory spraying, which can be further exploited to execute code. The file contents that are mostly used for such attacks are the ones related to images (such as \texttt{BMP, TIFF}) or fonts (such as \texttt{TFF}). In other cases, such as the direct execution of \texttt{EXE} payloads, the content is not necessarily malformed, but simply directly executed (this also avoid to execute malicious \texttt{JavaScript} code to exploit the vulnerability further). The execution of \texttt{EXE} payloads can also lead to the generation of additional, malicious payloads (such as \texttt{VBS}), whose final goal is dropping the final piece of malware.

\section{Machine Learning for PDF Malware Detection}
\label{sec:detection}

Machine learning is nowadays widely used to detect document-based infection vectors. In particular, concerning PDF files, multiple detectors were developed in the last decade that implemented such technology. Therefore, the aim of this Section is providing an overview of the characteristics of such detectors. However, as this survey focuses on the implications of adversarial attacks against machine learning systems, this Section only concerns systems that employ supervised machine learning to perform detection, meaning that we will not discuss PDF malware detectors that employ rule-based or non-supervised approaches (\eg ~\cite{tzermias11-eurosec,snow11-usenix,schmitt12-pst,vatanamu12-virology,willems12-acsac,lu13-hicss,liu14-dsn,xu17-usenix}). For a more detailed description of such systems, we refer the reader to more general purpose surveys~\cite{nissim15-cose,elingiusti18-chapter}. 

\begin{figure*}[thbp]
	\centering
	\includegraphics[width=0.9\textwidth]{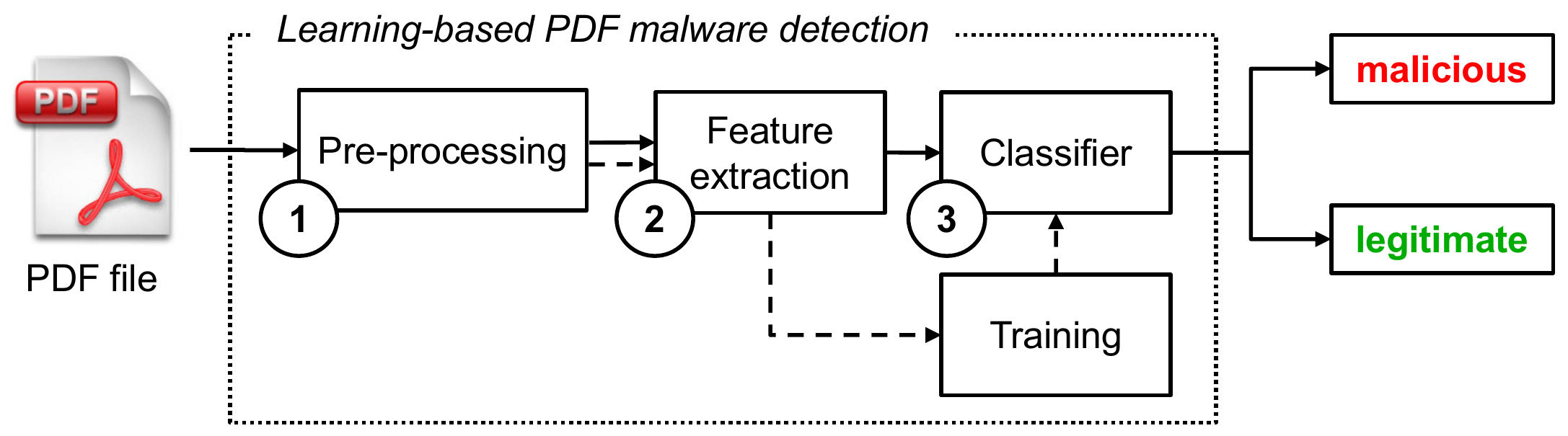}
	\caption{Graphical architecture of a learning-based PDF malware detection tool.}
	\label{sec:detection:fig:system-outline}
\end{figure*}

The primary goal of machine-learning detectors for malicious documents is discriminating between benign and malicious files. They can operate by analyzing and classifying information retrieved either from the structure or the content of the document. More specifically, all systems aimed to detect malicious PDF files share the same basic structure (reported in Figure~\ref{sec:detection:fig:system-outline}), which is composed of three main components~\cite{maiorca19-sp}:
\begin{enumerate}
	\item \textbf{Pre-Processing}. This component parses PDF files by isolating the data that are crucial for detection. For example, it can extract \texttt{JavaScript} or \texttt{ActionScript} code, select specific keywords or metadata, and so forth.
	\item \textbf{Feature Extraction}. This component operates on the information extracted during the pre-processing phase, by converting it to a vector of numbers. Such vector can represent, for example, the presence of specific keywords or API calls, or also the occurrences of certain elements in the file.
	\item \textbf{Classifier}. The proper learning algorithm. Its parameters are first tuned during the training phase to reduce overfitting and guarantee the highest flexibility against polymorphic variants.  
\end{enumerate}

According to the three components described above, Table ~\ref{sec:detection:tab:pdfDetectors} provides a general overview of machine learning-based PDF detectors that have been released since $2008$.

\begin{table*}[t]
	\centering
	\caption{An overview of the main characteristics of current PDF malware detectors.} 
	\label{sec:detection:tab:pdfDetectors}
	\resizebox{\columnwidth}{!}{
	\begin{tabular}{@{}lcccccc@{}}
		\toprule
		\multicolumn{1}{l}{\emph{Detector}} & \multicolumn{1}{c}{\emph{Tool}} & \multicolumn{1}{c}{\emph{Year}} &  \multicolumn{2}{c}{Pre-processing}  & Features & Classifier \\ \midrule
		\multicolumn{1}{l|}{\textbf{Shafiq~\etal}~\cite{shafiq08-dimva}} &
		N-Gram &
		\multicolumn{1}{l|}{$2008$} &
		\multicolumn{1}{c|}{Static} & Custom & RAW Bytes & Markov \\
		\multicolumn{1}{l|}{\textbf{Tabish ~\etal}  ~\cite{tabish09-sigkdd}} & N-Gram & \multicolumn{1}{l|}{$2009$} & \multicolumn{1}{c|}{Static} & Custom & RAW Bytes & Decision Trees \\
		\multicolumn{1}{l|}{\textbf{Cova ~\etal}~\cite{cova10-www}} & 
		Wepawet &
		\multicolumn{1}{l|}{$2010$} & \multicolumn{1}{c|}{Dynamic} & JSand & JS-based & Bayesian \\
		\multicolumn{1}{l|}{\textbf{Laskov and \Srndic}~\cite{laskov11-acsac}} &
		PJScan &
		\multicolumn{1}{l|}{$2011$} & \multicolumn{1}{c|}{Static} & Poppler & JS-based & SVM \\
		\multicolumn{1}{l|}{\textbf{Maiorca~\etal}~\cite{maiorca12-mldm}} &
		Slayer &
		\multicolumn{1}{l|}{$2012$} & \multicolumn{1}{c|}{Static} & PDFID & Structural & Random Forest \\
		\multicolumn{1}{l|}{\textbf{Smutz and Stavrou}~\cite{smutz12-acsac}} & PDFRate-v1 & \multicolumn{1}{l|}{$2012$} & \multicolumn{1}{c|}{Static} & Custom & Structural & Random Forest \\
		\multicolumn{1}{l|}{\textbf{\Srndic and Laskov }~\cite{srndic13-ndss,srndic16-eurasip}} & Hidost & \multicolumn{1}{l|}{$2013$} & \multicolumn{1}{c|}{Static} & Poppler & Structural & Random Forest \\
		\multicolumn{1}{l|}{\textbf{Corona ~\etal}~\cite{corona14-aisec}} &
		Lux0R &
		\multicolumn{1}{l|}{$2014$} & \multicolumn{1}{c|}{Dynamic} & PhoneyPDF & JS-based & Random Forest \\
		\multicolumn{1}{l|}{\textbf{Maiorca~\etal}~\cite{maiorca15-icissp}} & Slayer NEO & \multicolumn{1}{l|}{$2015$} & \multicolumn{1}{c|}{Static} & PeePDF+Origami & Structural & Adaboost \\
		\multicolumn{1}{l|}{\textbf{Smutz and Stavrou}~\cite{smutz16-ndss}} & PDFRate-v2 & \multicolumn{1}{l|}{$2016$} & \multicolumn{1}{c|}{Static} & Custom & Structural & Classifier Ensemble \\
		\bottomrule
	\end{tabular}
	}
\end{table*}

Note how each system employs a combination of different components types (\eg, a specific pre-processing module with a specific learning algorithm or feature extractor). Therefore, we structure the remaining of the Section by describing how each component can be characterized, along with its strengths and weaknesses (which will be crucial when such components are observed from the perspective of adversarial machine learning).

\subsection{Pre-processing}
\label{sec:detection:subsec:preproc}

As reported in the previous Sections, the pre-processing phase is crucial to select data that are used for detection. Table~\ref{sec:detection:tab:pdfDetectors} showed that there are two major types of pre-processing: \emph{static} and \emph{dynamic}. Static preprocessing analyzes the PDF file without executing it (or its contents). Dynamic pre-processing employs techniques such as sandboxing or instrumentation to execute either the PDF file or its 
\texttt{JavaScript} content. While static analysis is considerably faster and does not require many computational resources, dynamic analysis is much more effective at detecting obfuscated samples, meaning that certain obfuscation routines are almost impossible to be de-obfuscated automatically (in many cases there are multiple stages of obfuscation). As employing a sandboxed Adobe Reader is quite expensive computationally and not practical for the end-user, most detection systems rely on static parsing. 

When describing pre-processing tools, we generally divide them into two categories, which we list in the following: 
\begin{itemize}
	\item \textbf{Third-party Processing}. This category includes parsers that have been already developed and tested, and that do not directly depend on detection tools, thus being included as external modules in the system. The main advantage of employing third-party parsers is that they include many functionalities that are less prone (although not immune, as we will discuss more in detail later) to bugs. However, they can also embed unneeded functionalities and can be quite heavy computationally. 
	\item \textbf{Custom Processing}. This category includes parsers that have been written from scratch, in order to adapt to the information required from the detection system. As such parsers are tailored to the operations of the detectors, their functionalities are rather limited and much prone to bugs, as typically they have not been extensively tested. 
\end{itemize}

Table~\ref{sec:detection:tab:pdfDetectors} clearly shows that the favorite developers choice is relying on already existent tools, mostly because of their resilience to bugs and capabilities to adapt to most file variants.  For this reason, in the following, we provide a more extensive description of third-party parsers, which is summarized in Table~\ref{sec:detection:subsec:preproc:tab:parsers}.

\begin{table*}[t]
	\centering
	\caption{An overview of third-party parsers employed by machine learning-based PDF malware detectors. Each parser has been evaluated concerning three key components of the file. When the parser can completely analyze that specific component, we use the term \emph{Complete}; when only certain operations (in brackets) can be performed, we employ the term \emph{Partial}; finally, when no support is provided, we use the term \emph{None}.}  
	\label{sec:detection:subsec:preproc:tab:parsers}
	\begin{tabular}{@{}lcccc@{}}
		\toprule
		\multicolumn{1}{l}{\emph{Parser}} & \multicolumn{1}{c}{PDF Structure} & \multicolumn{1}{c}{JavaScript} & Embedded Files\\ \midrule
		\multicolumn{1}{l|}{\textbf{Origami}  ~\cite{origami}} & Complete & Partial (Code Analysis) & Complete \\
		\multicolumn{1}{l|}{\textbf{JSand}  ~\cite{cova10-www}} & None & Complete & Partial (Analysis) \\
		\multicolumn{1}{l|}{\textbf{PDFId}~\cite{pdfid}} & Partial (Key Analysis) & None & None \\
		\multicolumn{1}{l|}{\textbf{PeePDF} ~\cite{peepdf}} & Partial (Obj. Analysis) & Partial (Code Analysis) & Complete \\
		\multicolumn{1}{l|}{\textbf{PhoneyPDF}  ~\cite{phoneypdf}} & None & Complete & None \\
		\multicolumn{1}{l|}{\textbf{Poppler}  ~\cite{poppler}} & Complete & Partial (Extraction) & Complete \\
		
		\bottomrule
	\end{tabular}
\end{table*}

The Table is organized as follows. Each parser analyzes three main elements of the PDF file: the \emph{PDF structure}, the embedded \texttt{JavaScript} \emph{code} and the presence of \emph{embedded files} of any type (including further PDF files). Such elements are analyzed with three degrees of complexity: \emph{Complete}, \emph{Partial} or \emph{None}. The impact of such complexity degrees depends on the analyzed PDF element. In the following, we provide a more detailed description of each degree of complexity for each PDF element: 

\begin{itemize}
	\item \textbf{PDF Structure}. It refers to all elements of the PDF structure that are not related to embedded code, such as direct or indirect objects, metadata, and so on. When parsers completely support PDF Structure, it means that they can not only \emph{extract} and analyze object and metadata, but also that they can perform \emph{structural changes} to the file, such as object injection or metadata manipulation. When the support is partial, we typically refer to parsers that are only able to analyze objects, but not to manipulate them. When the support is set to None, it means that the PDF Structure is not analyzed at all. \texttt{Poppler}~\cite{poppler} and \texttt{Origami} are the only parsers that provide the possibility of properly injecting and manipulating content inside the file.
	\item \textbf{JavaScript}. It refers to the embedded \texttt{JavaScript} code inside the file. When the support to \texttt{JavaScript} is complete, it means that the code can be either statically and dynamically analyzed (to overcome de-obfuscation), for example through instrumentation. When the support is partial, it means that the code can be only statically analyzed (or even only extracted, as it happens in \texttt{Poppler}~\cite{poppler}), leading to some limitations when heavy obfuscation is employed. Finally, when no support is provided, the \texttt{JavaScript} code is not even extracted. \texttt{JSand} and \texttt{PhoneyPDF}~\cite{cova10-www,phoneypdf} are the only parsers that completely support \texttt{JavaScript} instrumentation and execution.
	\item \textbf{Embedded Files}. It refers to the capability of parsers to extract or inject embedded files (such as executable, office documents, or even other PDF files). When the support is complete, parsers can either extract or inject embedded files into the original PDF. When the support is partial, embedded files can only be extracted (or analyzed). Finally, when no support is provided, it means that embedded files cannot be extracted. \texttt{Origami}, \texttt{PeePDF} and \texttt{Poppler}~\cite{origami,peepdf,poppler} support extraction and analysis of embedded contents. 
\end{itemize}

From the description provided by Table~\ref{sec:detection:subsec:preproc:tab:parsers}, it can be inferred that no parsers can extract or manipulate all elements of the PDF file, although some of them allow for more functionalities. For this reason, the choice of the parser to be used is related to the type of information that is needed by the learning algorithm to perform its functionality. In the following, we provide a brief description of each third-party parser. 

\begin{itemize}
	\item \textbf{Origami}~\cite{origami}. This parser, entirely written in Ruby, allows users to navigate the object structure of PDF files, to craft malicious files by injecting code or other objects, to decompress and decrypt streams, and so forth. Moreover, it embeds popular information to recognize \texttt{JavaScript} API-based vulnerabilities (see Section~\ref{sec:malware:exploit}).
	\item \textbf{JSand}~\cite{cova10-www}. This parser was part of the \texttt{Wepawet} engine to perform dynamic analysis of PDF files. It could execute the embedded \texttt{JavaScript} code to extract API calls and de-obfuscate code.  Moreover, it could inspect embedded executables to reveal the presence of additional attacks. Unfortunately, the \texttt{Wepawet} service is currently not available; hence it is not possible to test \texttt{JSand} anymore.
	\item \textbf{PDFId}~\cite{pdfid}. This parser has been developed to extract the PDF \emph{name objects} (see Section~\ref{sec:malware}). It does not perform additional analysis on embedded code or files. 
	\item \textbf{PeePDF}~\cite{peepdf}. This parser, entirely written in Python, can perform a complete analysis of the PDF file structure (without being able to inject objects). It allows to inject and extract embedded files, and it provides a basic static analysis of \texttt{JavaScript} code.
	\item \textbf{PhoneyPDF}~\cite{phoneypdf}. This parser, entirely written in Python, performs dynamic analysis of embedded \texttt{JavaScript} code through instrumentation. More specifically, it emulates the execution of the \texttt{JavaScript} code embedded in a PDF file, in order to extract API calls that are strictly related to the PDF execution. This parser does not perform any structural analysis or embedding files extraction.
	\item \textbf{Poppler}~\cite{poppler}. \texttt{Poppler} is a C++ library that is used by popular, open-source software such as \texttt{X-Pdf} to render the contents of the file. For this reason, the library features complete support to PDF Structure parsing and managing, as well as \texttt{JavaScript} code extraction and injection of embedded files.
\end{itemize}  

Concerning \emph{custom parsers}, it is important to observe that we could not access the tools that adopted such parsers, as their source was not publicly released. Hence, we could only refer to what has been stated in the released papers~\cite{shafiq08-dimva,tabish09-sigkdd,smutz12-acsac,smutz16-ndss}. While raw bytes parsers ~\cite{shafiq08-dimva,tabish09-sigkdd} only focused on extracting byte sequences from the PDF, the parser adopted by \texttt{PDFRate} ~\cite{smutz12-acsac,smutz16-ndss} analyzed and extracted the object structure of the PDF file, with a particular focus on PDF metadata. However, the latter parser has been used as a test-bench for adversarial attacks, and researchers proved it could be easily evaded~\cite{srndic14-sp,carmony16-ndss} (see the next Sections).

\subsection{Feature Extraction}
\label{sec:detection:subsec:features}

\begin{table*}[t]
	\centering
	\caption{An overview of the feature types selected by each machine learning-based PDF detector. Each field of the table further specifies the type of feature employed by each detector. Worth noting, when a specific feature type is not used by the detector, we put a \texttt{x} on that field. }   
	\label{sec:detection:subsec:features:tab:feats}
	\resizebox{0.9\columnwidth}{!}{
	\begin{tabular}{@{}lcccccc@{}}
		\toprule
		\multicolumn{1}{l}{\emph{Detector}} & \multicolumn{1}{c}{\emph{Tool}} & \multicolumn{1}{c}{\emph{Year}} & \multicolumn{1}{c}{Structural} & \multicolumn{1}{c}{JS-Based} & \multicolumn{1}{c}{RAW Bytes}\\ \midrule
		\multicolumn{1}{l|}{\textbf{Shafiq ~\etal}~\cite{shafiq08-dimva}} & N-Gram & \multicolumn{1}{l|}{$2008$}  & x & x & N-Grams \\
		\multicolumn{1}{l|}{\textbf{Tabish ~\etal}~\cite{tabish09-sigkdd}} & N-Gram & \multicolumn{1}{l|}{$2009$} & x & x & N-Grams \\
		\multicolumn{1}{l|}{\textbf{Cova ~\etal}~\cite{cova10-www}} & Wepawet & \multicolumn{1}{l|}{$2010$} & x & Execution-Based & x \\
		\multicolumn{1}{l|}{\textbf{Laskov and \Srndic}~\cite{laskov11-acsac}} & PJScan & \multicolumn{1}{l|}{$2011$} & x & Lexical & x \\
		\multicolumn{1}{l|}{\textbf{Maiorca~\etal}~\cite{maiorca12-mldm}} & Slayer & \multicolumn{1}{l|}{$2012$} & Keywords & x & x \\
		\multicolumn{1}{l|}{\textbf{Smutz and Stavrou}~\cite{smutz12-acsac}} & PDFRate-v1 & \multicolumn{1}{l|}{$2012$} & Metadata & x & x \\
		\multicolumn{1}{l|}{\textbf{\Srndic and Laskov}~\cite{srndic13-ndss,srndic16-eurasip}} & Hidost & \multicolumn{1}{l|}{$2013$} & Key. Sequence & x & x \\
		\multicolumn{1}{l|}{\textbf{Corona ~\etal}~\cite{corona14-aisec}} & Lux0R & \multicolumn{1}{l|}{$2014$} & x & API-Based & x \\
		\multicolumn{1}{l|}{\textbf{Maiorca~\etal} ~\cite{maiorca15-icissp,maiorca15-chapter}} &  Slayer NEO & \multicolumn{1}{l|}{$2015$} & Keywords & API-Based & x \\
		\multicolumn{1}{l|}{\textbf{Smutz and Stavrou}~\cite{smutz16-ndss}} & PDFRate-v2 & \multicolumn{1}{l|}{$2016$} & Metadata & x & x \\

		\bottomrule
	\end{tabular}
	}
\end{table*}

Feature extraction is essential to PDF malware classification. In this phase, data obtained from the pre-processing phase are further parsed and transformed into vectors of numbers of a fixed size. Table ~\ref{sec:detection:subsec:features:tab:feats} provides an overview of the feature types that are used by each PDF detector. We can divide the employed feature types into three categories: 

\begin{itemize}
	\item \textbf{Structural}. These features are related to the PDF structure, and most concern the presence or the occurrence of specific keywords (name objects) in the file. Others include metadata or the presence/count of indirect objects or streams.
	\item \textbf{JS-Based}. These features are related to the structure of \texttt{JavaScript} code. Most of them concern lexical characteristics of the code (for example, the number of specific operators in the file), used API calls, or information obtained from the code behavior (\eg, when shellcodes are embedded in the attack).
	\item \textbf{Raw Bytes}. This category includes features that concern sequences of bytes taken as n-grams (\ie, groups of n-bytes, where n is typically a small integer).
\end{itemize}

Most PDF detectors only implement feature extraction methodologies that include only one type of feature. Byte-level features were among the first ones employed to solve the problem of PDF malware detection, and the very first works in the field mostly adopted them ~\cite{shafiq08-dimva,tabish09-sigkdd}. Typically, these features are represented by simple sequences of bytes taken in groups of n, where n is a very small integer. The reason for such a small number is because the feature space can explode quite easily. Using $1$-gram means a total of $256$ features, while a $2$-gram means $65536$ features. For this reason, this solution has not been considered very practical on standard machine learning models. Moreover, byte-level features do not typically convey explainable information on the file characteristics.

\texttt{JavaScript}-based features have been mostly adopted by \texttt{Wepawet,PJScan} and \texttt{Lux0R}~\cite{cova10-www,laskov11-acsac,corona14-aisec}. The general idea behind using these features is to isolate functions that perform dangerous operations, as well as detecting the presence of the obfuscated code that is typically associated with malicious behaviors. \texttt{Wepawet}~\cite{cova10-www} extracted information obtained from the code execution in an emulated environment. Such information is mostly related to code behavior, such as the type of parameters that are passed to specific methods, the sequences of suspicious API calls during execution, the number of bytes that are allocated for string operation (which may be a hint of heap spraying), and so forth. \texttt{PJScan} resorted to lexical information extracted from the \texttt{JavaScript} code itself, such as the count of specific operators (such as \texttt{+} or \texttt{( )}) that are known for being abused when obfuscated code. Moreover, it performs additional checks on the length of strings to point out the presence of suspicious exploiting techniques (such as buffer overflow or heap spraying). Finally, \texttt{Lux0R} exclusively operates on \texttt{JavaScript} API calls that belong to the Adobe Reader specifications. In particular, each call is evaluated only once (after being extracted during the pre-processing phase), leading to a binary feature vector of calls. While such features are excellent to analyze and detect attacks that carry \texttt{JavaScript} code, they cannot represent other types of attacks, such as the ones that involve the use of embedded files.

Structural approaches have been considerably used in recent years. The main idea, in this case, was trying to address all possible attacks reported in Section~\ref{sec:malware:exploit}, by using the most general approach possible. This idea also revolves around the concept that malware samples are structurally different from benign ones. For example, they typically feature fewer pages than benign samples, and the representation of the content is significantly scarcer.  The first approaches (adopted by \texttt{Slayer} and by its extension \texttt{Slayer NEO} - which also employed a very reduced number of JavaScript-based features related to known vulnerable APIs)~\cite{maiorca12-mldm,maiorca15-icissp,maiorca15-chapter} focused on counting the occurrences of specific keywords in the file. The keywords were regarded as relevant when they appeared at least once on enough files in the training set. \texttt{Hidost}~\cite{srndic13-ndss,srndic16-eurasip} evolved these approaches by employing sequences of keywords. More specifically, each feature was extracted by walking the PDF tree and evaluating how keywords were sequentially referred. The number of features was limited to $1000$, as using all the possible features would have led to an explosion of the algorithm in terms of complexity. Finally, \texttt{PDFRate}~\cite{smutz12-acsac,smutz16-ndss} focused on more general-purpose features that included, among the others, the number of indirect objects, the properties of the stream contents (\eg, the number of upper-case and lower-case characters) and so forth. Moreover, the approach also makes use of information obtained from metadata to identify suspicious behaviors (for example, when popular tools do not generate the PDF file). Albeit structural approaches proved to be effective at detecting even \texttt{ActionScript}-based attacks, they exhibit some limitations that will be better described in the next Sections.

\subsection{Learning and Classification}
\label{sec:detection:subsec:learning}

Feature extraction can be regarded as the process of mapping an input PDF file $\vct z \in \set Z$, being $\set Z$ the abstract space containing all PDF files, onto a vector space $\set X \subseteq \mathbb R^{\con d}$. In the feature space $\set X$, each file is  represented as a numerical  vector $\vct x$ of $d$ elements.
From a mathematical perspective, we can thus characterize the feature extraction process as a mapping function $\phi: \set Z \mapsto \set X$ that maps PDF files onto a vector space. 
Thanks to this abstraction, it is possible to use any learning algorithm to perform classification of PDF documents. All PDF malware detectors resort to \emph{supervised approaches}, \ie, they require the labels of the training samples to be known. More specifically, a learning algorithm is \emph{trained} to recognize a set of known training samples $\set D = (\vct x_{i}, y_{i})_{i=1}^{\con n}$, labeled as either legitimate ($y=-1$) or malicious ($y=+1$).\footnote{Without loss of generality, we assume here that the malicious (legitimate) class is assigned the positive (negative) label.} During this process, the parameters of the learning algorithm (if any) are typically set according to some given performance requirements.
After training, the learning algorithm provides a classification function $f : \set X \mapsto \mathbb R$ that can be used to assign a real-valued score to each input sample $\vct x$ at test time. Without loss of generality, we can assume here that $\vct x$ is classified as malicious (positive) if $f(\vct x) \geq 0$, and legitimate (negative) otherwise.
Notably, the appropriate learning algorithm is selected depending on the features that are used for classification and on the training data at hand. Decision trees have been used by most PDF malware detectors, and proved to be  very effective to address this problem \cite{tabish09-sigkdd,maiorca12-mldm,smutz12-acsac,srndic13-ndss,corona14-aisec,maiorca15-icissp,smutz16-ndss}. In particular, ensemble models such as Random Forests or Boosting showed very high accuracy under limited false positives. The underlying reason is that such classifiers well adapt to heterogeneous information, and in particular to discrete feature values such as counts. However, depending on the feature types, other solutions may be adopted. For example, \texttt{PJScan}~\cite{laskov11-acsac} resorts to Support Vector Machines (SVMs), \texttt{Wepawet}~\cite{cova10-www} to Bayesian classifiers, and  Shafiq~\etal~\cite{shafiq08-dimva} to Markov models (to deal with $n$-gram-based feature representations). Nevertheless, tree-based classifiers generally reported better performances at detecting PDF malware (see Section \ref{sec:detection:subsec:results}). 

\subsection{Detection Results}
\label{sec:detection:subsec:results}

In the following, Table~\ref{sec:detection:subsec:features:tab:results} presents the results attained on PDF malware detection by the most popular, machine learning-based systems in the wild. Notably, the goal of the Table is \emph{not} to show which system performs best but to provide indications on how such detectors generally cope with PDF attacks in the wild. Direct comparison among the performances attained by all systems would not be fair, considering that each system was trained and tested on different samples, with different training/test splits and classifiers. Therefore, in order to build a Table that is coherent and meaningful, we followed these guidelines:
\begin{itemize}
	\item We considered the results attained by systems that exclusively detected PDF malware. Therefore, we ruled out \cite{shafiq08-dimva,tabish09-sigkdd,cova10-www}, as their datasets also included other malware types besides PDF.
	\item Our Table includes the overall number of malicious and benign samples, the percentage of the dataset used for training the system, the True Positive (TP) rate attained in correspondence of the relative False Positive (FP) rate (TP@FP), and the relative F1 score. TP@FP is among the most used performance measure in malware analysis, and it is very useful to indicate the performances of the systems at specific FP values. Keeping a low FP rate guarantees proper system usability (too many false alarms would disrupt the overall user experience).  Note that we referred to values that were \emph{explicitly} stated in each \emph{original} work (we did not consider any cross-analysis).
	\item Notably, many papers adopted significantly imbalanced datasets in their analysis. For this reason, we used F1 as an overall measure that considers the distribution of the data as a crucial element to measure performance. We point out that systems with higher F1-score are not necessarily better than the others. We included this evaluation to provide the reader with another perspective from which evaluating the results.   
\end{itemize}

Notably, these choices have been driven by the heterogeneous nature of the examined works. In particular, many of them did not report precise numbers of employed benign and malicious test samples, but only the overall test-set percentage concerning a precise number of malicious and benign samples. For this reason, when calculating the F1 scores for each system, we assumed that the test percentages were equally applied for benign and malicious samples. 

Some of the examined works reported multiple results attained by changing the features, classification parameters, and data distribution~\cite{laskov11-acsac,smutz12-acsac,srndic13-ndss,srndic16-eurasip}. Hence, for each system we focused on the following information (the reader may check the original works for more details):
\begin{itemize}
	\item \textbf{PJScan}~\cite{laskov11-acsac}. We considered as malicious the files that were regarded as \emph{detected} in the original paper, and as benign the files that were regarded as \emph{undetected}. We reported the performances obtained with native tokens and on \texttt{JavaScript} files only, as \texttt{PJScan} does not make any analysis of non-\texttt{JavaScript} files. 
	\item \textbf{PDFRate-v1}~\cite{smutz12-acsac}. We reported the performances related to the lowest number of false positives that are stated in the original work. 
	\item \textbf{Hidost}~\cite{srndic13-ndss,srndic16-eurasip}. We reported the result attained by the work in \cite{srndic13-ndss}, as it clearly states the attained TP and FP percentages. 
	\item \textbf{PDFRate-v2}~\cite{smutz16-ndss}. As performances were stated in the paper by considering multiple thresholds for classification, we report the results attained at 0.5 threshold. The choice was made to obtain false positives values that were similar to the ones attained by the other systems.
\end{itemize}

\begin{table*}[t]
	\centering
	\caption{Results attained at PDF malware detection by the most popular, machine learning-based detectors. The Table reports the overall malicious and benign samples, the training set percentages, the True Positives (TP) at False Positives (FP) rate, and the overall F1 score.}   
	\label{sec:detection:subsec:features:tab:results}
	\resizebox{\columnwidth}{!}{%
	\begin{tabular}{@{}lccccccc@{}}
		\toprule
		\multicolumn{1}{l}{\emph{Detector}} & \multicolumn{1}{c}{\emph{Tool}} & \multicolumn{1}{c}{\emph{Year}} & \multicolumn{1}{c}{Mal. Samples} & \multicolumn{1}{c}{Ben. Samples} &
		\multicolumn{1}{c}{Train. Perc. (\%)} & \multicolumn{1}{c}{TP@FP (\%)} &
		\multicolumn{1}{c}{F1} \\ \midrule
		\multicolumn{1}{l|}{\textbf{Laskov and \Srndic}~\cite{laskov11-acsac}} & PJScan & \multicolumn{1}{l|}{$2011$} & 15,279 & 960 & 50 & 71.94@16.35 & 0.832 \\
		\multicolumn{1}{l|}{\textbf{Maiorca~\etal}~\cite{maiorca12-mldm}} & Slayer & \multicolumn{1}{l|}{$2012$} & 11,157 & 9989 & 57 & 99.5@0.02 & 0.989\\
		\multicolumn{1}{l|}{\textbf{Smutz and Stavrou}~\cite{smutz12-acsac}} & PDFRate-v1 & \multicolumn{1}{l|}{$2012$} & 5,297 & 104,793 & 9.1 & 93.27@0.02 & 0.801 \\
		\multicolumn{1}{l|}{\textbf{\Srndic and Laskov }~\cite{srndic13-ndss,srndic16-eurasip}} & Hidost & \multicolumn{1}{l|}{$2013$} & 82.142 & 576,621 & 33.7 & 99.73@0.06 & 0.825 \\
		\multicolumn{1}{l|}{\textbf{Corona~\etal}~\cite{corona14-aisec}} & Lux0R & \multicolumn{1}{l|}{$2014$} & 12,548 & 5,234 & 70 & 99.27@0.05 & 0.986 \\
		\multicolumn{1}{l|}{\textbf{Maiorca~\etal} ~\cite{maiorca15-icissp,maiorca15-chapter}} &  Slayer NEO & \multicolumn{1}{l|}{$2015$} & 11,138 & 9,890 & 57 & 99.81@0.07 & 0.969 \\
		\multicolumn{1}{l|}{\textbf{Smutz and Stavrou}~\cite{smutz16-ndss}} & PDFRate-v2 & \multicolumn{1}{l|}{$2016$} & 5,297 & 104,793 & 9.1 & 99.5@0.05 & 0.667\\
		\bottomrule
	\end{tabular}
	}
\end{table*}

The attained results show some interesting trends. First and foremost, almost all systems attained very high accuracy at detecting PDF malware with low false positives. Such positive results mean that various information (feature) types can be equally effective at detecting PDF malware.  

Second, there is a consistent F1-score difference between \texttt{Slayer}~\cite{maiorca12-mldm} and its evolved \texttt{Slayer NEO}~\cite{maiorca15-icissp} version, and between \texttt{PDFRate-v1}~\cite{smutz12-acsac} and \texttt{PDFRate-v2}~\cite{smutz16-ndss}, where the attained F1-scores decreased in the most recent versions of the tools. In particular, we observe that this decrease is due to a higher number of false positives. Notably, this effect is particularly evident on \texttt{PDFRate-v2}, where imbalanced datasets significantly penalize the F1 score when false positives increase. Finally, we observe that other works used most of the dataset for training \cite{maiorca12-mldm,corona14-aisec,maiorca15-icissp,maiorca15-chapter}, which means that such systems would require more training data to perform classification correctly. 

\section{Adversarial Attacks against PDF Malware Detectors}
\label{sec:adversarial}

In this Section, we start by formalizing a threat model for PDF malware detection, inspired from previous work in the area of adversarial machine learning, and then we will use it to categorize existing and potentially-new adversarial attacks against such systems.

As highlighted in a recent survey~\cite{biggio18-pr}, the first attempts to formalize adversarial attacks against learning algorithms date back to the decade 2004-2014~\cite{dalvi04-kdd,lowd05-kdd,barreno06-asiaccs,globerson06-icml,barreno10,huang11,kloft12-jmlr,biggio12-icml,biggio14-tkde,biggio14-svm-chapter,biggio14-ijprai}, prior to the recent discovery of \emph{adversarial examples} against deep neural networks~\cite{szegedy14-iclr,goodfellow15-iclr}. 
PDF malware has provided one of the major case studies in the literature of adversarial machine learning over these years, as its inherent structure allows for fine-grained modifications that well-adapt to how adversarial attacks are typically performed.
As we will discuss in the remainder of this paper, this is true in particular for evasion attacks, \ie, attacks in which PDF malware is manipulated at test time to evade detection.
Preliminary attempts in crafting evasive attacks against PDF malware detectors were first described by Smutz and Stavrou~\cite{smutz12-acsac}, and subsequently by \Srndic and Laskov~\cite{srndic13-ndss}, even though they were based on heuristic strategies that were able to mislead linear classification algorithms successfully.
To our knowledge, the very first work proposing optimization-based evasion attacks against PDF malware detectors is the work by Biggio~\etal~\cite{biggio13-ecml}. In that work, the authors were able to show that even nonlinear models were vulnerable to \emph{adversarial} PDF malware, conversely to what envisioned in~\cite{srndic13-ndss}. The attack was done by selecting the feasible manipulations to be performed on each malware sample via a gradient-based optimization procedure, in which the attacker aimed to minimize the classifier's prediction confidence on the malicious class (\ie, to get maximum-confidence predictions on the opposite class, namely, the benign one).
Worth remarking, the gradient-based procedure described in that paper has been the first to demonstrate the vulnerability of learning algorithms to optimization-based evasion attacks, even before the discovery of adversarial examples against deep networks~\cite{szegedy14-iclr,goodfellow15-iclr}.

\subsection{Threat Modeling and Categorization of Adversarial Attacks}
\label{sec:adversarial:subsec:model}

The threat model proposed in this Section aims to provide a unified treatment of both attacks developed in the area of adversarial machine learning and attacks developed specifically against PDF malware detectors. As we will see, this does not only enable us to categorize existing attacks under a consistent framework, clarifying the (sometimes implicit) assumptions behind each of them. The proposed threat model will also help us to identify new potential attacks that may threaten PDF malware detectors in the near future, as well as novel research directions for improving the security of such systems, inspired by recent findings in the area of adversarial machine learning.

Leveraging the taxonomy of adversarial attacks initially provided by Barreno~\etal~\cite{barreno06-asiaccs,barreno10}, and subsequently expanded by Huang~\etal~\cite{huang11}, Biggio and Roli~\cite{biggio18-pr} have recently provided a unified threat model that categorizes adversarial attacks based on defining the \emph{attacker's goal}, her \emph{knowledge} of the target system, and her \emph{capability} of manipulating the input data.\footnote{We refer to the attacker as feminine here due to the popular role impersonated by Eve (or Carol) in cryptography.}
In the following, we discuss how to adapt this threat model to the specific case of PDF malware detectors.

\subsubsection{Attacker's Goal} 
\label{sec:adversarial:subsec:model:subsubsec:goal}

The attacker's goal is defined based on the desired security violation. When speaking about systems' security, an attacker can cause \textbf{integrity}, \textbf{availability} or even \textbf{privacy} violations. Violating system integrity means having malware samples undetected, without compromising normal system operation for legitimate users. An example of integrity violation is when benign PDF files are injected with a payload that is undetected by anti-malware engines. In this way, the user still visualizes the file content, but other operations still occur in the background. 
Availability is compromised when many benign (but also malware) samples are misclassified, causing a denial of service for legitimate users effectively. This situation may occur on PDF files when the attacker injects multiple scripting (benign) instructions to trigger multiple fake alerts due to the presence of codes (which cannot be, alone, indicators of maliciousness).
Finally, privacy is violated if the system may leak confidential information about its users, the classification model used, and even the training data used to learn it. For example, an attacker can incrementally change some characteristics of the PDF file (for example, by adding text, using fonts, adding scripting codes, and so forth) and send them to the target classifier to see how it reacts to such changes.

\subsubsection{Attacker's Knowledge}
\label{sec:adversarial:subsec:model:subsubsec:knowledge}

The attacker can have different levels of knowledge of the targeted system, including: ($i$) the training data $\set D$, consisting of $n$ training samples and their labels, \ie, $(\vct x_i, y_i)^n$; ($ii$) the feature set $\set X \subseteq \mathbb R^d$, which is strongly related to the PDF detector components depicted in Figure~\ref{sec:detection:fig:system-outline} and Table~\ref{sec:detection:tab:pdfDetectors} (see Section~\ref{sec:detection}). More specifically, the attacker may know which pre-processing and feature extraction algorithms are employed, along with the extracted feature types; ($iii$) the classification function $f : \set X \mapsto \mathbb R$ (to be compared against zero for classification), along with the objective function minimized during training (if any), its hyperparameters and, potentially, even the classifier parameters learned after training.
The attacker's knowledge can be conceptually represented in an abstract space $\Theta$, whose elements correspond to the aforementioned components ($i$)-($iii$) as $\vct \theta=(\set D, \set X, f)$.
Depending on the assumptions made on ($i$)-($iii$), there may be different attack scenarios, described in the following, and compactly summarized in Table~\ref{sec:adversarial:subsec:model:subsubsec:knowledge:tab:taxonomy}.

\textbf{Perfect-Knowledge (PK) White-Box Attacks.} In this scenario, we assume that the attacker knows everything about the target system, \ie, $\vct \theta_{\rm PK}=(\set D, \set X, f)$. Even though this may rarely occur in practical settings, this scenario is useful to provide an upper bound on the performance degradation incurred by the system under attack, and to understand how far from the worst case more realistic evaluations are.

\textbf{Limited-Knowledge (LK) Gray-Box Attacks.} This category of attacks, in general, assume that the attacker knows the feature representation $\set X$, but she does not have full knowledge of the training data $\set D$ and the classification function $f$. In particular, it is often assumed that the attacker can collect a surrogate data set $\hat{\set D}$ resembling that used to train the target classifier from a similar source (\eg, a public repository).\footnote{We use here the \emph{hat} symbol to denote limited knowledge of a given component.}
Regarding the classification function $f$, the attacker may know the type of learning algorithm used (\eg, the fact that the classifier is a linear SVM) and, ideally, its hyperparameters (\eg, the value of the regularization parameter $C$ used to learn it), although this may not be strictly required. However, the attacker is assumed not to know the exact classifier's trained parameters (\eg, the weights of the linear SVM after training), but she can potentially get feedback from the classifier about its decisions and labels.
Under these assumptions, the attacker can estimate the classification function from $\hat{\set D}$, by training a \emph{surrogate classifier} $\hat f$. We thus denote this attack scenario with $\vct \theta_{\rm LK}=(\hat{\set D}, \set X, \hat{f})$.
Notably, these attacks may also include the case in which the attacker knows the trained classifier $f$ (white-box attacks), but optimizing the attack samples against the target function $f$ may be not effective. The reason is that, for certain configurations of the classifier's hyperparameters, the objective function in the white-box case may become too noisy and difficult to optimize with gradient-based procedures, due to the presence of many poor local minima or null gradients (a phenomenon known as \emph{gradient obfuscation}~\cite{athalye18,papernot17-asiaccs}). Therefore, it is preferable for the attacker to optimize the attack samples against a surrogate classifier with a smoother objective function and then test them against the target one. This is a common procedure also used to evaluate the \emph{transferability} of attacks~\cite{biggio13-ecml,papernot17-asiaccs,dong18-cvpr,liu17-iclr,demontis18-arxiv}.

\begin{table*}[t]
	\centering
	\caption{An overview of the knowledge levels held by an attacker, according to the elements of the knowledge space. When a component of the space is fully known, we represent it with a checkmark (\checkmark), while we use a \texttt{x} when the element is partially known or unknown.}  
	\label{sec:adversarial:subsec:model:subsubsec:knowledge:tab:taxonomy}
	\begin{tabular}{@{}lccc@{}}
	\toprule
	\multicolumn{1}{l}{\emph{Knowledge Level}} & \multicolumn{1}{c}{$\set D$} & \multicolumn{1}{c}{$\set X$} &
	\multicolumn{1}{c}{$f$} \\ \midrule
	\multicolumn{1}{l|}{\textbf{White box (PK)}}  & \checkmark & \checkmark & \checkmark  \\
	\multicolumn{1}{l|}{\textbf{Gray box (LK)}} & \texttt{x} & \checkmark & \texttt{x}   \\
\multicolumn{1}{l|}{\textbf{Black box (ZK)}} & \texttt{x} & \texttt{x} & \texttt{x}   \\
	
	\bottomrule
\end{tabular}
\end{table*}

\myparagraph{Zero-Knowledge (ZK) Black-Box Attacks} The term \emph{zero knowledge} is typically used in literature to indicate the possibility that the attacker can query the system to obtain feedback on the labels and the provided score, and optimize the attack samples in a black-box fashion~\cite{tramer16-usenix,xu16-ndss,papernot17-asiaccs,chen17-aisec,dang17-ccs}. However, it should be clarified that the attacker still has some minimal knowledge of the system. For example, she knows that the classifier has been designed to perform specific tasks (\eg, detecting PDF files), and she has an idea of what kind of transformations should be made on the samples to attempt evasion. Hence, some details of the feature representation are still known (\eg, the employed feature types, whether static or dynamic analysis is performed, etc.). The same considerations can be done about knowledge of the training data; \eg, it is obvious that a system designed to recognize specific PDF malware has been trained with benign and malicious PDF files. 
Thus, in agreement with Biggio and Roli~\cite{biggio18-pr}, we characterize this setting as $\vct \theta_{\rm ZK} = (\hat{\set D}, \hat{\set X}, \hat{f})$. Note that using surrogate classifiers is not strictly necessary here~\cite{tramer16-usenix,xu16-ndss,dang17-ccs,chen17-aisec}; however, it is possible to learn a surrogate classifier (potentially on a different feature representation) and check whether the crafted attack samples \emph{transfer} to the target classifier. Feedback from classifier's decisions on specifically-crafted query samples can be also used to refine and update the surrogate model~\cite{papernot17-asiaccs}.

\subsubsection{Attacker's Capability} 
\label{sec:adversarial:subsec:model:subsubsec:capability}
The attacker's capability of manipulating the input data is defined in terms of the so-called \emph{attack influence}, as well as by some data manipulation constraints. The attack influence defines whether the attacker can only manipulate \textbf{test data} (\emph{exploratory} influence), or also the \textbf{training data} (\emph{causative} influence). The latter is possible, \eg, if the system is retrained online using data collected during operation, which can be manipulated by the attacker~\cite{barreno06-asiaccs,huang11,biggio14-tkde,biggio18-pr}.
Depending on whether the attack is staged at \emph{training} or \emph{test} time, different data manipulation constraints can be defined.
These define the changes that can be concretely performed by the attacker to evade the system.
For example, to evade malware detection at test time, malicious code must be modified without compromising its intrusive functionality. This strategy may be employed against systems based on static code analysis, by injecting instructions or code that will never be executed~\cite{srndic14,biggio13-ecml,demontis17-tdsc,grosse17-esorics}.
For training-time attacks, instead, the constraints typically impose that the attacker can only control a small fraction of the training set~\cite{barreno06-asiaccs,huang11,kloft12-jmlr,biggio18-pr,jagielski18-sp}.
In both cases, as discussed in~\cite{biggio18-pr}, the attacker's capability can be formalized in terms of mathematical constraints along with the optimization problem defined to craft the attack samples.

\subsubsection{Summary of Attacks}
\label{sec:adversarial:subsec:model:subsubsec:summary}

We describe here the potential attacks that can be perpetrated against machine-learning algorithms, according to the assumptions made on the attacker's goal and on her capability of manipulating the input data~\cite{biggio18-pr}.
Table~\ref{sec:adversarial:subsec:model:subsubsec:summary:tab:taxonomy} introduces the set of adversarial attacks that have been considered to date. Notably, each of these attacks can be performed according to different levels of knowledge, as described in Section~\ref{sec:adversarial:subsec:model:subsubsec:knowledge}.

\begin{table}[htbp]
\centering
	\caption{An overview of adversarial attacks against learning systems, adapted from~\cite{biggio18-pr}.}  
	\label{sec:adversarial:subsec:model:subsubsec:summary:tab:taxonomy}
\resizebox{.85\textwidth}{!}{%
\begin{tabular}{p{3.5cm}p{3.5cm}p{3.5cm}p{3.8cm}}
\toprule
 & \multicolumn{3}{c}{\textbf{Attacker's Goal}} \\ 
\textbf{Attacker's Capability} & \multicolumn{1}{c}{\textit{Integrity}} & \multicolumn{1}{c}{\textit{Availability}} & \multicolumn{1}{c}{\textit{Privacy}} \\
\midrule \midrule
Test data &  \begin{tabular}{l}
     Evasion  
\end{tabular} &
\begin{tabular}{c} - \end{tabular} &
\begin{tabular}{l}
Model Extraction/Stealing\\
Model Inversion\\
Membership Inference
\end{tabular}  \\
\midrule
Training data &
\begin{tabular}{l}
Poisoning (Integrity) \\
Backdoor
\end{tabular}
  & 
  \begin{tabular}{l}
  Poisoning (Availability)
  \end{tabular}
    & \begin{tabular}{c} - \end{tabular}  \\ 
\bottomrule
\end{tabular}
}
\end{table}

\myparagraph{Evasion Attacks} In this setting, the attacker attempts to manipulate test samples to have them misclassified as desired~\cite{biggio13-ecml}. Evasion attacks are also commonly referred to as \emph{adversarial examples}, especially when staged against deep-learning algorithms for image classification~\cite{szegedy14-iclr,biggio18-pr}. These attacks exploit specific weaknesses of a previously-trained model, and they have been widely used to show test-time vulnerabilities of learning-based malware detectors. 

\myparagraph{Poisoning Attacks} Poisoning attacks target the training stage of the classifier. The attacker intentionally injects wrongly-labeled samples into the training set, aiming to decrease the detection capabilities of the classifier. 
If the attack aims to indiscriminately increase the classification error at test time, causing a denial of service to legitimate system users, it is referred to as a \emph{poisoning availability} attack~\cite{barreno06-asiaccs,barreno10,huang11,mei15-aaai,biggio12-spr,biggio13-icb,biggio18-pr,biggio17-aisec,jagielski18-sp}.
Conversely, if the attack is targeted to only have few samples misclassified at test time, then it is named as a \emph{poisoning integrity} attack~\cite{barreno06-asiaccs,barreno10,huang11,biggio12-spr,biggio13-icb,kloft12-jmlr,biggio18-pr,biggio17-aisec,jagielski18-sp,koh17-icml}. We also include \emph{backdoor attacks} into this category, as they also aim to cause specific misclassifications at test time by compromising the training process of the learning algorithm~\cite{yujie18-ccs,liu18-ndss,gu17,chen17}. 
Their underlying idea is to compromise a model during the training phase or, more generally, at design time (this may include, \eg, also modifications to the architecture of a deep network by addition of specific layers or neurons), with the goal of enforcing the model to classify \emph{backdoored} samples at test time as desired by the attacker. Backdoored samples may include malware files with a specific signature or images with a specific subset of pixels set to given values.
Once the model recognizes this specific signature (\ie, the backdoor), it should output the desired classification. A popular example is the stop sign with a yellow sticker attached on it (\ie, the signature used to activate the backdoor), which is misclassified as a speed limit sign by the backdoored deep network considered in~\cite{gu17}.
The overall intuition is that these vulnerable models can then be released to the public, to be used as open-source pre-trained models in other open-source tools or even commercial products, and consequently make the latter also vulnerable to the same backdoor attacks.

\myparagraph{Privacy Attacks} Privacy-related attacks aim to steal information about unknown models for which the attacker is given black-box query access. In these attacks, the attacker sends specific test samples against the target model, with the aim of obtaining information about: ($i$) the model itself, via \emph{model extraction} (or \emph{model stealing}) attacks~\cite{tramer16-usenix}; or  ($ii$) the data used to train it, via \emph{model inversion} attacks (to reconstruct the training samples)~\cite{fredrikson15-ccs}  or \emph{membership inference} attacks (to evaluate whether a given sample has been used to train the target model or not)~\cite{shokri17-sp,nasr18-ccs}.

In the remainder of this manuscript, we provide detailed insights into each attack concerning the problem of PDF malware detection.

\subsection{Evasion Attacks against PDF Malware Detectors} \label{sec:adversarial:subsec:evasion}

Research work in attacking PDF malware detectors focused mostly on evasion attacks~\cite{smutz12-acsac,srndic13-ndss,biggio13-ecml,biggio14-svm-chapter}. 
As highlighted by our previous categorization of adversarial attacks, evasion attacks aim to violate system \emph{integrity} by manipulating the input PDF file structure at \emph{test time}. 

In the specific case of attacks against PDF malware detectors, we can identify two main families of evasion attacks, \ie, \emph{optimization-based}~\cite{biggio13-ecml,biggio14-svm-chapter,srndic14-sp,xu16-ndss} and \emph{heuristic-based}~\cite{smutz12-acsac,srndic13-ndss,maiorca13-asiaccs,corona14-aisec,maiorca15-icissp,carmony16-ndss,smutz16-ndss,maiorca19-sp} attacks.
Optimization-based evasion attacks perform fine-grained modifications to the input sample in order to minimize (maximize) the probability that the sample is classified as malicious (benign).
Heuristic-based evasion attacks aim also to create evasive samples, but they are not formulated in terms of an optimization problem. They rather provide reasonable modifications that are expected to cause misclassifications of malware samples at test time, including, \eg, trying to mimic the structure of benign files.
Within these two broad categories of attacks, we will then specify whether each attack is carried out under a white-box, gray-box or black-box attack scenario, to highlight the set of assumptions made on the attacker's knowledge, as encompassed by our threat model. We will also discuss the practical difficulties associated to concretely staging the attack against a real system, which are inherent to the creation of the evasive (or \emph{adversarial}) PDF malware samples.

This issue has been widely known in the literature of adversarial machine learning, especially for optimization-based attacks, under the name of \emph{inverse feature-mapping problem}~\cite{huang11,biggio13-ecml,biggio14-tkde,biggio14-ijprai}. This problem arises from the fact that most of the optimization-based attacks proposed thus far craft evasive instances only in the  \emph{feature space}, \ie, by directly manipulating the values of the feature vector $\vct x$, without creating the corresponding evasive PDF malware sample. The modifications are typically constrained to make it possible to create the actual PDF file, at least in principle; however, how to invert the representation $\vct x$ to obtain the corresponding PDF file as $\vct z = \phi^{-1}(\vct x)$ has only been rarely considered. 
This point is particularly of interest, and goes beyond the problem of adversarial PDF malware; for example, adversarial attacks manipulating Android and binary malware are subject to the same issues~\cite{demontis17-tdsc,kolosnjaji18-eusipco}.
The problem here amounts to creating a malware sample exhibiting the desired feature vector, modified to evade the target system, which is not always an easy task. For example, in the case of PDF malware, injecting material may change the file \emph{semantics} (\ie, the file may behave differently than the original one) or even break the malicious functionality of the embedded exploitation code. 

We summarize the attacks proposed thus far to craft \emph{adversarial PDF malware} in Table~\ref{sec:adversarial:subsec:evasion:subsec:category:tab:attacks}. Our taxonomy is organized along four axes: ($i$) the family of evasion attacks (\ie, either heuristic-based or optimization-based); ($ii$) the attacker's knowledge of the learning model (\ie, whether she has white-box, gray-box or black-box access to the target classifier); ($iii$) the target system(s); and ($iv$) whether the attack has been staged at the \emph{input level} (\ie, by creating the adversarial PDF malware sample) or at the \emph{feature level} (\ie, by only modifying the  feature values of the attack samples, without creating the corresponding PDF files). 

We provide below a more detailed description of the attacks that have been performed against PDF malware detectors, following the proposed categorization (Sections~\ref{sec:adversarial:subsec:evasion:subsubsec:opt}-\ref{sec:adversarial:subsec:evasion:subsubsec:heur}).
We then discuss some insights on how to tackle the inverse feature-mapping problem and create adversarial PDF malware samples. As we will see, this can be achieved by injecting content into PDF files through the exploitation of vulnerabilities in their parsing mechanisms (Section~\ref{sec:adversarial:subsec:evasion:subsubsec:practical}). 

\begin{table*}[t]
	\caption{An overview of the evasion attacks proposed thus far against PDF malware detectors. This taxonomy considers the attack family (\ie, optimization- or heuristic-based), the attacker's knowledge of the target system (\ie, whether she has white-box, gray-box, or black-box access), the target system(s), and whether the attack has been staged at the input level (\ie, creating the adversarial PDF malware samples) or at the feature level (\ie, only manipulating their feature values, without actually creating the PDF files).}   
	\label{sec:adversarial:subsec:evasion:subsec:category:tab:attacks}
	\resizebox{\columnwidth}{!}{
	\begin{tabular}{@{}lcccccc@{}}
		\toprule
		\multicolumn{1}{l}{\emph{Work}} & \multicolumn{1}{l}{\emph{Year}} & \multicolumn{1}{c}{Heur./Opt.} &
		\multicolumn{1}{c}{Knowl. (B/G/W)} & \multicolumn{1}{c}{Target System(s)} &  \multicolumn{1}{c}{Real Sample}\\ \midrule
		\multicolumn{1}{l|}{\textbf{Smutz and Stavrou}  \cite{smutz12-acsac}} & \multicolumn{1}{l|}{$2012$}  & Heur. & G & PDFRate-v1 & No \\
		\multicolumn{1}{l|}{\textbf{\Srndic and Laskov} \cite{srndic13-ndss}} & \multicolumn{1}{l|}{$2013$} &  Heur. & B,W & Hidost & No \\
		\multicolumn{1}{l|}{\textbf{Biggio ~\etal}  \cite{biggio13-ecml,biggio14-svm-chapter}} & \multicolumn{1}{l|}{$2013$} & Opt. & G,W & Slayer & No \\
		\multicolumn{1}{l|}{\textbf{Maiorca~\etal}  \cite{maiorca13-asiaccs}} & \multicolumn{1}{l|}{$2013$} & Heur. & B & Wepawet, PDFRate-v1, PJScan & Yes \\
		\multicolumn{1}{l|}{\textbf{Corona ~\etal} \cite{corona14-aisec}} & \multicolumn{1}{l|}{$2014$} & Heur. & B & Lux0R & No \\
		\multicolumn{1}{l|}{\textbf{\Srndic and Laskov} \cite{srndic14-sp}} & \multicolumn{1}{l|}{$2014$}  & Opt. & W & PDFRate & Yes \\
		\multicolumn{1}{l|}{\textbf{Maiorca~\etal}  \cite{maiorca15-icissp,maiorca15-chapter}} & \multicolumn{1}{l|}{$2015$} & Heur. & B & Wepawet, PDFRate-v1, PJScan, Slayer N. & Yes \\
		\multicolumn{1}{l|}{\textbf{Carmony~\etal} \cite{carmony16-ndss}} & \multicolumn{1}{l|}{$2016$} & Heur. & B & PDFRate-v1, PJScan & Yes \\
		\multicolumn{1}{l|}{\textbf{Xu~\etal} \cite{xu16-ndss}} & \multicolumn{1}{l|}{$2016$} & Opt. & B & PDFRate-v1, Hidost & Yes \\
		\multicolumn{1}{l|}{\textbf{Smutz and Stavrou} \cite{smutz16-ndss}} & \multicolumn{1}{l|}{$2016$} & Heur., Opt. & B,W & PDFRate-v2 & Yes \\
		\multicolumn{1}{l|}{\textbf{Maiorca and Biggio} \cite{maiorca19-sp}} & \multicolumn{1}{l|}{$2019$} & Heur. & B & PDFRate-v1, PJScan, Slayer N., Hidost & Yes \\
		\bottomrule
	\end{tabular}
	}
\end{table*}

\subsubsection{Optimization-based Evasion Attacks} 
\label{sec:adversarial:subsec:evasion:subsubsec:opt}

Evasion attacks (also known as \emph{adversarial examples}) consist of minimizing the classifier's score on the malicious class $f(\vct x^\prime)$ with respect to the input adversarial sample $\vct x^\prime$~\cite{biggio13-ecml}. The changes to the input vector $\vct x^\prime$ are constrained to reflect feasible manipulations on the source malicious PDF file $\vct x = \phi(\vct z)$. In the case of PDF malware detection, they are typically restricted to only consider the injection of content.
The problem of optimizing adversarial PDF malware samples has been originally formulated by Biggio~\etal~\cite{biggio13-ecml} as:
\begin{eqnarray}
\label{eq:opt1} \argmin_{\vct x^\prime} && f(\vct x^\prime) - \lambda  g(\vct x^\prime) \, , \\
\label{eq:opt2} {\rm s.t.} && \vct x \preceq \vct x^\prime \, , \, {\rm and} \, \| \vct x^\prime - \vct x \|_1 \leq \varepsilon \, ,
\end{eqnarray}
where the first constraint $\vct x \preceq \vct x^\prime$ reflects content injection (\ie, each component of $\vct x^\prime$ has to be greater or equal than the corresponding one in $\vct x$), and the second one bounds the number of injected elements to be not greater than $\varepsilon$ (as it essentially counts in how many elements $\vct x$ and $\vct x^\prime$ are different).
The function $g(\vct x^\prime)$ reflects how similar the manipulated sample $\vct x^\prime$ is to the benign distribution, and $\lambda$ is a trade-off parameter.  
This trick was introduced in Biggio~\etal~\cite{biggio13-ecml,biggio14-svm-chapter} to facilitate evasion by avoiding poor local minima of $f(\vct x^\prime)$ (which do not typically allow the attack algorithm to find a correct evasion point).
In fact, when $\lambda$ is small, the attacker may find an evasion point by typically injecting very few objects, and such point is normally quite different from both the malicious and benign training samples. When $\lambda$ increases, the attack point is modified in a more significant manner, but it also becomes harder to distinguish its manipulated feature representation to that of benign samples.
The aforementioned optimization problem is usually solved by projected gradient descent~\cite{biggio13-ecml,biggio14-svm-chapter}, and it has been used in follow-up work to generate adversarial PDF malware~\cite{srndic14-sp,smutz16-ndss}, adversarial binaries~\cite{kolosnjaji18-eusipco} and adversarial images against deep networks~\cite{melis17-vipar}.
With respect to the area of adversarial images against deep networks, many different algorithms have been proposed to generate the so-called \emph{adversarial examples}~\cite{szegedy14-iclr,goodfellow15-iclr,papernot16-sp,carlini17-sp,madry18-iclr}. They are nevertheless all based on the idea of formulating the attack as an optimization problem, and solve it using different gradient-based optimization algorithms.
We refer the reader to~\cite{biggio18-pr} for a comprehensive survey on this topic. 

To solve the aforementioned problem with gradient-based optimization, the attacker needs to be able to estimate the gradient of the objective function, which in principle requires white-box access to the target classifier $f(\vct x^\prime)$, and ideally also to the training data (to estimate the gradient of $g(\vct x^\prime)$). However, it has been shown in~\cite{biggio13-ecml} that such gradient can be estimated reliably even if the attacker has only gray-box access to the target system. The underlying idea is to first collect a surrogate training set and query the target system to relabel it (\eg, if the target system is provided as an online service).\footnote{However, it has been shown that if the surrogate data is well representative of the training data used by the target system, the relabeling procedure is not even required~\cite{biggio13-ecml,demontis18-arxiv}.} 
A surrogate learner $\hat f$ can then be trained on such data, to approximate the target function $f$, while an estimate $\hat g$ of the benign sample distribution $g$ can be obtained directly from the surrogate data.
Attacks can be thus staged against the surrogate learner and then \emph{transferred} to the target system. Within this simple trick, one can craft gradient-based attacks with both white-box and gray-box access to the target learner.\footnote{Despite follow-up work has defined these attacks as black-box transfer attacks~\cite{papernot17-asiaccs}, we prefer to name them gray-box attacks, as such attacks implicitly assume that the attacker knows at least the feature representation.}
In the context of PDF malware, white-box and gray-box gradient-based attacks have been used in~\cite{biggio13-ecml,biggio14-svm-chapter,srndic14-sp,smutz16-ndss}. Black-box attacks have been more recently proposed in~\cite{xu16-ndss}, based on the idea of minimizing the value of $f(\vct x^\prime)$ by only querying the target classifier in an iterative manner through using a genetic algorithm.
These three families of attack are described below. We also discuss how they can be seen as specific instances of the aforementioned optimization problem.

\myparagraph{White-Box Gradient-based Evasion Attacks} This category includes the gradient-based attacks proposed by Biggio~\etal~\cite{biggio13-ecml,biggio14-svm-chapter} and by \Srndic~\etal~\cite{srndic14-sp}, assuming white-box access to the target model (which includes knowledge of its internal, trained parameters).
Biggio~\etal~\cite{biggio13-ecml,biggio14-svm-chapter} focused on evading \texttt{Slayer}~\cite{maiorca12-mldm}, and considering three different learning algorithms: SVMs with the linear and the RBF kernel, and neural networks. Their results showed that, even with a minimal number of injected keywords (less than $10$ against the linear SVM), it is possible to evade PDF malware detection.
These results showed, for the first time, that non-linear malware detectors can be vulnerable to test-time attacks (conversely to what had been previously reported in~\cite{srndic13-ndss}).
Despite the constrained optimization problem in Eqs.~\eqref{eq:opt1}-\eqref{eq:opt2} allows in principe the creation of the corresponding PDF malware samples, this was not concretely demonstrated in~\cite{biggio13-ecml} (as the analysis was only limited to the manipulation of numerical feature values).

To overcome this limitation, \Srndic and Laskov~\cite{srndic14-sp} expanded the work in~\cite{biggio13-ecml} by performing a practical evasion of \texttt{PDFRate-v1}~\cite{smutz12-acsac}. They considered the same optimization problem with white-box access to the target system, but injected the selected features for evasion directly into the malicious PDF file, after the \texttt{EOF} tag. Albeit effective at evading the target system, such an injection strategy may be easily countered by improving the parsing process~\cite{srndic14-sp}. Notably, as \texttt{PDFRate-v1} employs standard decision trees (which are non-differentiable), the authors replaced the classifier with a surrogate, differentiable SVM. Even with this characteristic, the attack can be considered as white-box, as the attacker possesses complete knowledge of the target system.
More recently, Smutz~\etal~\cite{smutz16-ndss} tested the same attack against an evolved version of \texttt{PDFRate} (which we refer to as \texttt{PDFRate-v2}), which adopted a customized ensemble of trees as a classifier. However, due to the non-differentiable nature of the trees, the authors directly replaced them with an ensemble of SVMs. Notably, the attack was performed directly against the SVM ensemble, which was not used as a surrogate. The attained results showed that gradient-based attacks completely evaded the system. 

\myparagraph{Gray-Box Gradient-based Evasion Attacks} This scenario has been explored by Biggio~\etal~\cite{biggio13-ecml,biggio14-svm-chapter} and by \Srndic and Laskov~\cite{srndic14-sp}. We point out here its importance, as it shows what happens when gradient-descent based attacks are performed under more realistic scenarios. Notably, it would be unfeasible for the attacker to have perfect knowledge of the target system (including the classifier's trained parameters). For this reason, as it is easily predictable, the efficacy of gradient-descent attacks is reduced by the fact that the attacker has to learn a surrogate algorithm (and to use surrogate data) to mount the attack. However, experiments performed against \texttt{Slayer} showed that all classifiers were completely evaded by simply increasing the number of changes to the file. Most were evaded after $30-40$ changes, which means injecting that number of keywords. This attack showed that, even under realistic scenarios, gradient descent can be a very effective approach to evade PDF malware detectors.

\myparagraph{Black-Box GA-based Evasion Attacks} This attack has been proposed by Xu~\etal~\cite{xu16-ndss}. 
The underlying idea is to optimize the objective function in Eq.~\eqref{eq:opt1} (with $\lambda=0$, \ie, to only minimize $f(\vct x^\prime)$), using a genetic algorithm (GA) that iteratively queries the target system in a black-box scenario. 
In this case, the input PDF malware sample is directly modified by injecting and deleting features (accordingly, the constraint $\vct x \preceq \vct x^\prime$ in Eq.~\eqref{eq:opt2} is not considered here), 
and then dynamic analysis is used to verify that its malicious functionality has been preserved.
This attack has been staged against \texttt{PDFRate-v1} and \texttt{Hidost}, and proven to be very effective at evading both systems. 
While this attack is the only one that considers removal of objects from PDF files, it is also very computationally demanding. The underlying reason is twofold: ($i$) the  genetic algorithm used in the optimization process requires performing a large number of queries (as it does not exploit any structured information like gradients to drive the search in the optimization space), and ($ii$) the attack requires running dynamic analysis on the manipulated samples. While this may not be an issue for an attacker whose goal is only developing successful attacks, GA-based evasion may not be suitable for generating attacks under limited time, computational resources and, most importantly, number of queries to the target system.

\subsubsection{Heuristic-based Evasion Attacks}
\label{sec:adversarial:subsec:evasion:subsubsec:heur}

Heuristic-based evasion attacks had been largely used before optimization-based attacks were found to be more effective. They include strategies that do not directly optimize an objective function, and this is why they are typically less effective than optimization-based attacks.
Heuristic-based attacks attempt evasion by making malicious samples more similar to benign ones. This effect is typically obtained either by adding benign features to malicious files (\emph{mimicry}) or by injecting malicious payload in benign files (\emph{reverse mimicry}). According to the knowledge possessed by the attacker, we distinguish among the three evasion strategies discussed below.

\myparagraph{White-Box Mimicry} The main idea of this attack is manipulating malware samples to make them as close as possible to the benign data (in terms of their conditional probability distribution or distance in feature space), under white-box access to the trained model, \ie, by injecting the most discriminant features for the target classifier directly into the input sample. This implementation was adopted in~\cite{smutz12-acsac} to bypass \texttt{PDFRate-v1}. The attained results showed that it was possible to decrease the classifier accuracy by more than $20\%$ by only injecting the top 6 discriminant features in each malware sample. A similar approach was adopted in~\cite{srndic13-ndss}, where the authors tested the robustness of \texttt{Hidost} by injecting the features that would influence the classifier decision the most. The attained result showed that it was possible to evade the tree-based classifiers with high probability, but this was not the case for nonlinear SVMs with the RBF kernel.  

\myparagraph{Black-Box Mimicry} In this strategy, the attacker injects into a malicious sample all the objects contained in a benign file, aiming to subvert the classifier decision. To this end, the attacker only needs to have black-box access to the target classifier and to collect some benign PDF files.
The easiest way to perform this attack is to blindly copy the entire content of a benign PDF file into the malicious sample. This approach was adopted by Corona~\etal~\cite{corona14-aisec} to test the robustness of \texttt{Lux0R}. In particular, the authors added all the JavaScript API-based features belonging to multiple benign samples to the given malicious PDF files and measured how the classifier detection was affected. They also measured the changes in detection rate as the number of injected files increased. Results showed that, due to the dynamic nature of the features employed, the classifier was still able to detect most of the malicious samples in the wild, despite the modifications they received. \Srndic and Laskov~\cite{srndic13-ndss} also adopted a similar approach to verify the robustness of \texttt{Hidost} (trained, in this case, with an RBF SVM - a different case to the tests they made with white-box mimicry against tree classifiers), showing that their approach was robust against this kind of mimicry.

\begin{figure}[t]
	\centering
	\includegraphics[width=.95\textwidth]{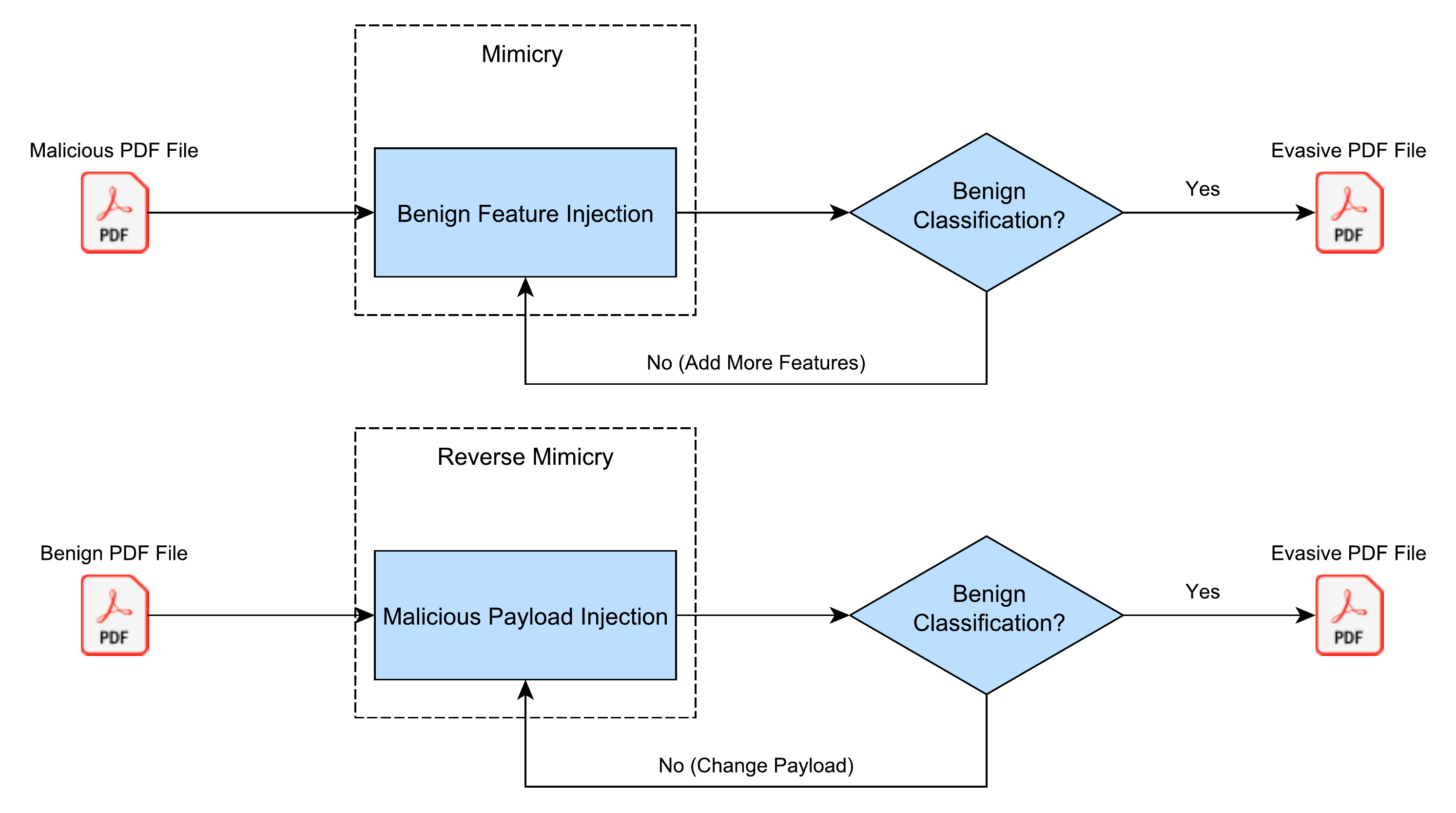}
	\caption{Conceptual flow of \emph{mimicry} (top) and \emph{reverse-mimicry} (bottom) attacks.}
	\label{sec:adversarial:subsec:evasion:subsubsec:heur:fig:mimicry}
\end{figure}

\myparagraph{Black-Box Reverse Mimicry} In this attack, the attacker injects a specifically-crafted, malicious payload into a file that is recognized as benign by the classifier, by only exploiting black-box access to the classifier. The idea is that the corresponding feature modifications should not be sufficient to change the classification of the source sample from benign to malicious~\cite{maiorca13-asiaccs,maiorca15-icissp,smutz16-ndss,maiorca19-sp}. Such a strategy is exactly the opposite of mimicry attacks: while the latter inject benign information into a malicious file, reverse mimicry tends to minimize the amount of malicious information injected (by changing the injected payload when the attack fails). This difference is depicted in the flow diagrams of Figure~\ref{sec:adversarial:subsec:evasion:subsubsec:heur:fig:mimicry}.

There are three types of reverse-mimicry attacks, according to different types of malicious information that can be injected:
\begin{itemize}
	\item \emph{JS Embedding}. This approach injects \texttt{JavaScript} codes directly inside the file objects. Current implementations feature the injection of only one \texttt{JavaScript} code, but it is technically possible to scatter the code through multiple objects in the file. Notably, this would increase the probability of the attack to be detected, as more keywords have to be used to indicate the presence of the embedded code.
	
	\item \emph{EXE Embedding}. This technique exploits the \texttt{CVE-2010-1240} vulnerability, thanks to which it is possible to run an executable from the PDF file itself directly. In the implementation proposed by Maiorca~\etal~\cite{maiorca13-asiaccs}, the EXE payload was injected through the creation of a different version. 
	\item \emph{PDF Embedding}. This strategy features the injection of malicious PDF files in the objects of the target PDF file. More specifically, attackers can inject specific keywords that cause the embedded file to open automatically. This technique can be particularly effective, especially because multiple embedding layers can be easily created (for example, embedding a PDF file in another PDF file, which is finally embedded in another file). In~\cite{maiorca15-icissp}, \texttt{PeePDF} was employed to carry out such embedding strategy. However, the embedding process is prone to bugs, due to the complexity of the PDF format. Nevertheless, it is possible to employ libraries such as \texttt{Poppler} to improve the injection process.
\end{itemize}

The efficacy of reverse-mimicry attacks has been explored in various works. Maiorca~\etal~\cite{maiorca13-asiaccs}. demonstrated that all reverse mimicry variants were able to evade systems that adopted structural features, such as \texttt{PDFRate-v1}. Further works~\cite{maiorca15-icissp,maiorca15-chapter,smutz16-ndss} proposed possible strategies to mitigate such attacks, which will be described more in detail in Section~\ref{sec:discussion}. 

\begin{figure}[t]
	\centering
	\includegraphics[width=.95\textwidth]{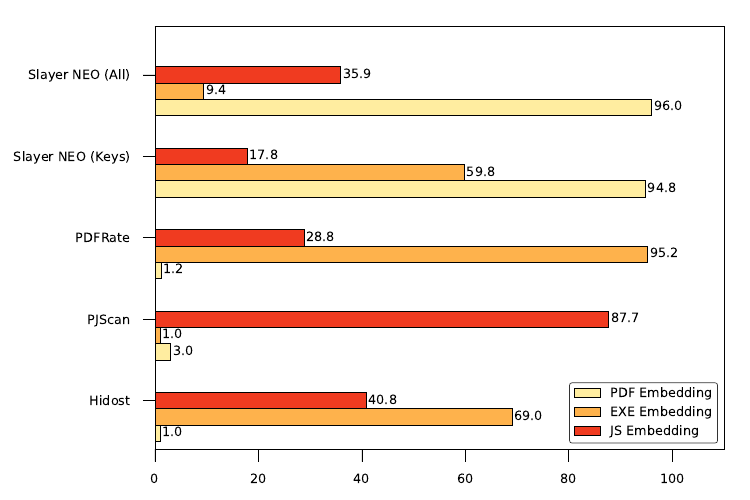}
	\caption{Results attained by detection systems against reverse mimicry attacks. Each system has been tested against $500$ samples for each attack variant ($1500$ samples in total). Results are reported by showing the percentage of detected samples.}
	\label{sec:adversarial:subsec:evasion:subsubsec:heur:fig:rev_results}
\end{figure}

We now summarize the results attained by testing evasive examples created with reverse-mimicry strategies. In particular, we report the results attained by \cite{maiorca19-sp}, in which $500$ samples for each reverse mimicry variant (for a total of $1500$ samples) were created and tested against multiple systems in the wild \cite{laskov11-acsac,smutz12-acsac,srndic13-ndss,smutz16-ndss,maiorca15-chapter}. Such comparison is the most recent and fair between multiple systems on evasive datasets. All systems (except for \texttt{PDFRate-v1}, which was provided as an online service\footnote{\texttt{PDFRate-v2} was not available during the experiemnts in~\cite{maiorca19-sp}.}) were trained with the same dataset composed by more than $20000$ malicious and benign files in the wild. Figure \ref{sec:adversarial:subsec:evasion:subsubsec:heur:fig:rev_results} reports the attained results. For each attack and system, we report the percentage of evasive samples that were detected. Note that \texttt{Slayer NEO} has been tested in two variants: the \emph{keys} variant reflects the work made in~\cite{maiorca12-mldm} (in which the system only operated by extracting keywords as features), while the \emph{all} variant reflects the full functionality of the system, as tested in~\cite{maiorca15-chapter}.

Results clearly show that each system has its strengths and weaknesses. For example, \texttt{Slayer NEO} well performs against PDF embedding attacks due to its capability of extracting and analyzing embedded files, while \texttt{PJscan} is excellent at detecting \texttt{JS} embedding attacks and \texttt{PDFRate-v1} provides reliable detection of \texttt{EXE} embedding. However, none of the tested systems can effectively detect all reverse-mimicry attacks.

\myparagraph{Other Black-Box Attacks} Other black-box attacks involve empirical attempts to exploit vulnerabilities of the components that belong to the target detectors, such as their parsers. This concept has been explored on a broader way by Carmony~\etal~\cite{carmony16-ndss}, who showed that each parser (including advanced ones) on which PDF detectors are based could be evaded due to implementation bugs, design errors, omissions, and ambiguities. For this reason, attackers can be motivated to mostly target the pre-processing modules of detection systems, thus efficiently exploiting their vulnerabilities. Indeed, the authors created working proofs of concept with which they were able to evade both third-party and custom (\texttt{PDFRate}) parsers.

\subsubsection{Practical Implementation}
\label{sec:adversarial:subsec:evasion:subsubsec:practical}
As already mentioned in Section ~\ref{sec:adversarial:subsec:evasion} and in Table~\ref{sec:adversarial:subsec:evasion:subsec:category:tab:attacks}, a critical problem of adversarial attacks is the creation of the real attack samples starting from the evasive feature vectors (\ie, the so-called \emph{inverse feature-mapping} problem). More specifically, the goal of the attacker is injecting/removing information into/from the PDF file while keeping its semantics intact (\ie, the file should exactly work as the unaltered version). However, due to the object-based structure of the PDF format (see Section~\ref{sec:malware}), manipulating each feature value independently may not always be feasible.
The corresponding file manipulations may compromise its malicious functionality, especially if the attack also requires removing content.

\begin{figure}[t]
	\centering
	\includegraphics[width=0.7\textwidth]{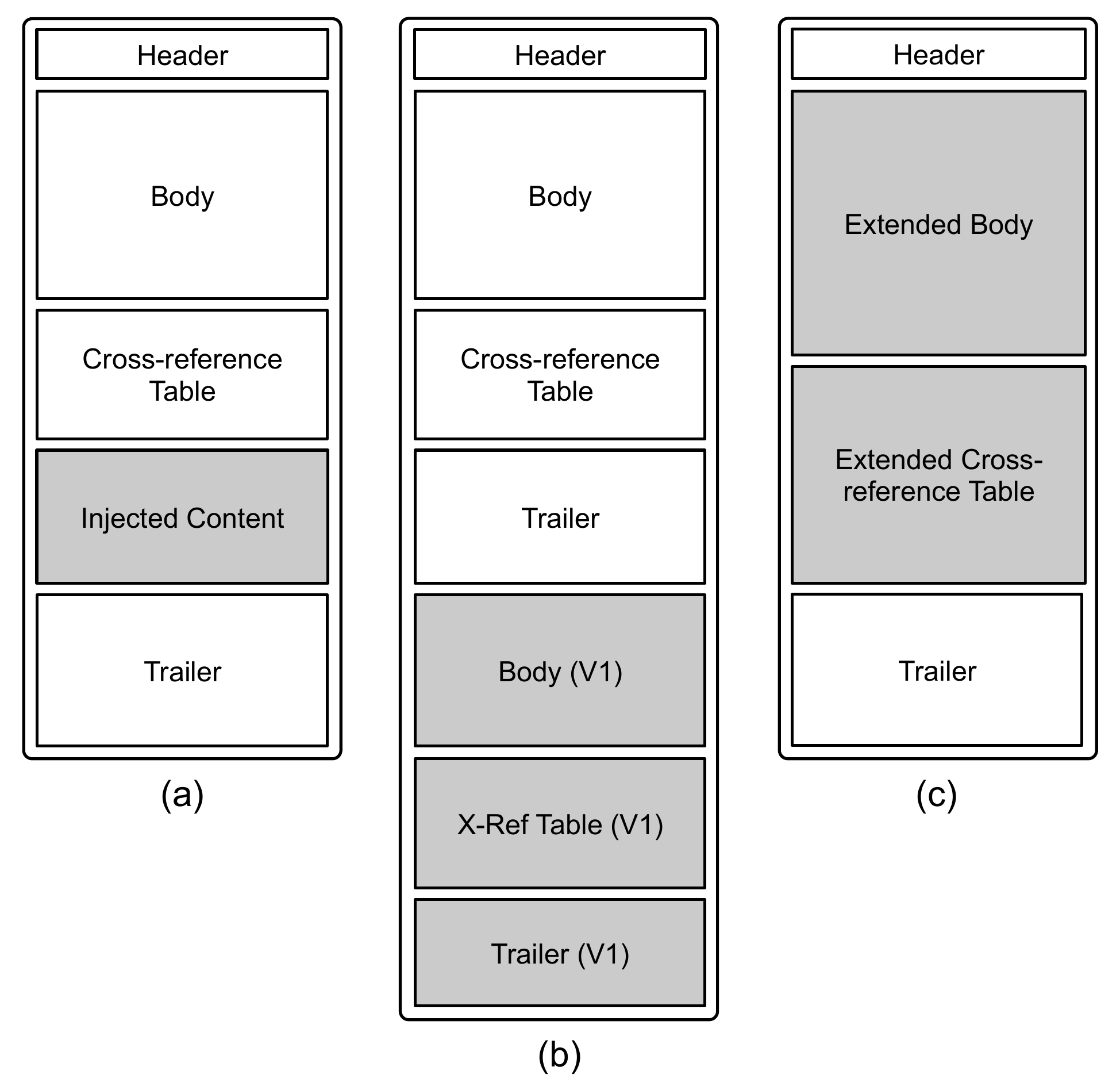}
	\caption{Content injection in PDF files, performed according to three possible strategies: (a) Injection after X-Ref; (b) Versioning; and (c) Body Injection.}
	\label{sec:adversarial:subsec:evasion:subsubsec:practical:fig:inj}
\end{figure}

Table~\ref{sec:adversarial:subsec:evasion:subsec:category:tab:attacks} showed which works implemented the real attack samples either by addressing the inverse feature-mapping problem or by directly manipulating the PDF malware sample. In both cases, three major strategies were used to inject material while minimizing the risk of compromising the intrusive functionality of PDF malware, as depicted in Figure~\ref{sec:adversarial:subsec:evasion:subsubsec:practical:fig:inj}~\cite{maiorca19-sp}. We now provide a comprehensive description of each technique, by also referring to the works where they have been employed. 
Worth noting, these strategies can also be exploited to address the inverse feature-mapping problem and implement concrete attacks at the input level from the feature-level attack strategies proposed in~\cite{smutz12-acsac,srndic13-ndss,biggio13-ecml,biggio14-svm-chapter,corona14-aisec}.

\myparagraph{(a) Injection after X-Ref} In this strategy, objects are injected after the X-Ref table, and in particular after the EOF marker of the file. These objects are never parsed by Adobe Reader or similar, as their presence is not marked in the cross-reference table. This strategy is easy to implement, but it strongly relies on exploiting vulnerabilities of the parsing process. Like clearly explained by Carmony~\etal~\cite{carmony16-ndss}, all PDF detector parsers suffer from vulnerabilities, as none of them fully implements the PDF specifications released by Adobe. Such weaknesses are particularly evident in custom parsers, which may hide some strong misalignments to the behavior of Adobe Reader. However, such injection strategy can be made ineffective by merely patching the pre-processing module of the PDF malware detector to be consistent with Adobe Reader. Injection after X-Ref has been mostly employed by \Srndic and Laskov~\cite{srndic14-sp} to evade \texttt{PDFRate-v1}, which employs a customized parser. Albeit easy to counteract, this strategy has been used to simplify the injection of different types of structural features (for example, upper/lower-case characters). The same strategy has been employed by Smutz~\etal ~\cite{smutz16-ndss} for their experiments. 

\myparagraph{(b) Versioning} In this strategy, attackers use the \emph{versioning} mechanism of the PDF file format, \ie, injecting a new body, X-Ref table, and trailer, as the user directly modified the file (\eg, by using an external tool - see Section~\ref{sec:malware}. This strategy is more advanced than the previous one, as the Cross-Reference table parses the injected objects, and therefore they are considered as legitimate by the Reader itself. In this way, it is also straightforward to add embedded files or other malicious (or benign) contents that can be used as an aid to perpetuate a more effective attack. Such injection strategy can, however, be countered by extracting the content of each version of the PDF file and process it separately. Maiorca~\etal~\cite{maiorca13-asiaccs,maiorca15-icissp,maiorca15-chapter,maiorca19-sp} employed this strategy to generate EXE Embedding attacks (belonging to reverse mimicry). More specifically, they leveraged the \texttt{Metasploit} framework~\cite{metasploit} to automatically generate the infected evasive samples. In this way, they showed that the versioning mechanism could be easily automatized by off-the-shelf tools to generate evasive variants. 

\myparagraph{(c) Body Injection} In this strategy, attackers directly operate on the existing PDF graph, adding new objects to the file body and re-arranging the X-Ref table accordingly. This strategy is more complicated to implement and to detect, as it seamlessly adds objects in a PDF file by reconstructing the objects and their position on the X-Ref Table.  In this way, it is possible to manage and re-arrange the X-Ref table objects without corrupting the file~\cite{carmony16-ndss}. Existing objects can also be modified by adding other name objects and rearranging the X-Ref table positions accordingly. Notably, it is essential to ensure that the embedded content (\ie, the exploitation code) is correctly executed when the merged PDF is opened. The correct execution of the embedded content is often not easy to achieve, as it requires injecting additional objects specifically for this purpose. This strategy was employed by Maiorca~\etal~\cite{maiorca13-asiaccs,maiorca15-icissp,maiorca15-chapter,maiorca19-sp} (and also used by Smutz~\etal~\cite{smutz16-ndss} and Carmony~\etal~\cite{carmony16-ndss}) to generate JS and PDF embedding attacks. In this way, the attacker can exactly choose in which part of the file body the malicious payload can be injected, thus making the automatic detection of such attacks harder. 

An extension of the aforementioned body-injection strategy has been adopted by Xu~\etal~\cite{xu16-ndss}, to account also for object removal and replacement. In particular, after each manipulation of the file body (including object addition, removal or replacement), the evasive sample is automatically tested on a a Cuckoo sandbox. If the attempted change disrupts the functionality of the file (\ie, the PDF malware does not contact a malicious target URL anymore), the original file is restored. Albeit computationally expensive, this strategy allows one to precisely verify whether it is possible to perform such specific changes to the source PDF file.

\subsection{Poisoning and Privacy Attacks against PDF Malware Detectors}
\label{sec:adversarial:subsec:poisoningprivacy}

In this Section, we discuss two other popular categories of attack defined in the literature of adversarial machine learning, under the name of poisoning and privacy attacks~\cite{huang11,biggio18-pr}.
Even though, to our knowledge, such attacks have never been considered in the context of PDF malware detection, we discuss here how they can be used to pose new threats to PDF malware detectors.

\subsubsection{Poisoning Attacks} 

Poisoning attacks aim to reduce the detection capabilities of PDF detectors by injecting wrongly-labeled samples in the classifier training set. According to the taxonomy proposed in Section~\ref{sec:adversarial:subsec:model:subsubsec:summary}, we distinguish between three possible attacks:

\myparagraph{Poisoning Availability Attacks}  This attack can be carried out against online services that ask for the user feedback about the classification results (such as \texttt{PDFRate-v1}, which used to be online available). More specifically, the attacker can submit several malicious PDF files for the analysis. When the system asks for feedback, she can intentionally claim that such samples are benign, hoping that the system gets retrained by including the wrongly-labeled samples. If the attacker has no explicit control on the labeling process, she may construct benign samples which contain spurious malicious features, and malicious samples with benign content, in a way that preserves their original labeling. Similar attacks have been staged against anti-spam filters~\cite{nelson08,huang11}, to eventually increase the classification error at test time, and cause a denial of service to legitimate users.

\myparagraph{Poisoning Integrity Attacks} This attack is similar to the one employed in poisoning availability. However, instead of aiming to increase the classification error at test time to cause a denial of service, the goal of the attacker here is to cause the misclassification of a specific subset of malicious PDF files at test time. For example, given a PDF malware sample which is correctly recognized by the target system, the attacker may start injecting benign samples with spurious malicious characteristics extracted from the malicious PDF file into the training set of the target classifier. Once updated, the target classifier may thus no longer correctly detect the given PDF malware sample.
Overall, poisoning integrity attacks aim to facilitate evasion at test time.

\myparagraph{Backdoor Attacks}  The goal of this attack is the same as poisoning integrity, \ie, to facilitate evasion at test time, even though the attack strategy may be implemented in a different manner. In particular, the attacker may publicly release the backdoored model, which may be subsequently used in some publicly-available online services or commercial tools. If this happens, the attacker can craft her malicious samples (including the backdoor activation signature) to bypass detection.

\subsubsection{Privacy Attacks} Privacy attacks in the context of PDF malware detection may aim to primarily steal information about the classification model or the training data. They can be organized into three categories:

\myparagraph{Model Stealing/Extraction Attacks} This attack can be performed on systems whose classification scores (or other information) are available~\cite{tramer16-usenix}. The attacker can submit PDF files with progressive modifications to their structure (or scripting code) to infer which changes occur to the information retrieved from the models (in a similar fashion to what performed by~\cite{xu16-ndss}). In this way, the attacker may eventually reconstruct the detection model with high accuracy, and sell an alternative online service at a cheaper price.

\myparagraph{Model Inversion Attacks} The attack strategy is similar to model stealing attacks, but the goal is reconstructing the employed training data~\cite{fredrikson15-ccs}. This may violate user privacy if the attacker is able to reconstruct some benign PDF sample containing private information.

\myparagraph{Membership Inference Attacks} This attack strategy is similar to the previous two cases, but the goal is to understand if a specific sample was used as part of the training data. For example, this can be useful to infer if the system was trained with PDF files that can be easily found on search engines, or if the system resorts to data available from publicly-distributed malware datasets~\cite{shokri17-sp,nasr18-ccs}.

\section{Open Issues and Research Challenges}
\label{sec:discussion}

We discuss here the current open issues related to PDF malware detection and  sketch some promising future research directions.
More specifically, further research can be developed from two perspectives, both suggested from the research area of adversarial machine learning: ($i$) demonstrating novel \emph{adversarial attacks} and potential threats against PDF malware detectors, and ($ii$) improving their \emph{robustness} under attack, by leveraging previous work on defensive mechanisms from adversarial machine learning. In the following, we discuss both perspectives.

\subsection{Adversarial Attacks} Considering what we pointed out in Sections~\ref{sec:adversarial:subsec:evasion} and~\ref{sec:adversarial:subsec:poisoningprivacy}, it is evident that research on attacks against learning-based PDF malware detectors mainly focused on evasion. In particular, state-of-the-art work has been carried out by following two main directions: on the one hand, the creation of concrete, working adversarial examples by using empirical approaches, with the drawback of increasing the probability of failing the attacks due to too limited knowledge of the target system; on the other hand, the development of evasion algorithms, leading to approaches that allowed very efficient evasion without the creation of real samples (due to, for example, changes that could not be correctly performed in practice, such as deleting specific keywords). Notably, one critical problem to be solved is related to inverse feature mapping, \ie, creating real samples from the corresponding evasive feature vectors. Normally, to preserve the original behavior of the file, injecting information is typically safer than removing it. Nevertheless, such an action could compromise the overall file semantics and visualization properties. This effect could be triggered, \eg, by embedding font- or pages-related keywords in specific objects. 

Recent work has demonstrated that it is possible to remove specific structural features (\eg, keywords) from PDF files without compromising their functionality.~\cite{xu16-ndss}. However, there is still a lot of space for research on this topic. More specifically, attackers could focus on identifying a set of features that can be safely deleted (depending on the file context), or even replaced with equivalent ones. For example, one could remove every reference to \texttt{JavaScript} code and replace them with another language such as \texttt{ActionScript}. Concerning this aspect, it would also be intriguing to inspect the dependence between certain features (for example, deleting one keyword may force the attacker also to delete a second one). In this way, one may think of increasing the efficacy of gradient-descent algorithms by including a selection of erasable/replaceable information.

Finally, alternatives to evasion attacks can be explored. As stated in Section \ref{sec:adversarial:subsec:poisoningprivacy}, poisoning and privacy attacks have yet to be carried out against PDF learners. One particularly interesting aspect is how such attacks can be useful in practice against current detection systems. For example, poisoning PDF detectors may compromise the performances of publicly available (or even open source) learning models models.  Likewise, model-stealing strategies can be employed to extract the characteristics of unknown detectors, leading to the development of more effective evasion attacks.

\subsection{Robust Malware Detection} 
From what we pointed out in Sections~\ref{sec:detection} and~\ref{sec:adversarial}, it is clear that every released detector features specific weaknesses that are either related to its parser, feature extractor, or classifier. However, while most state-of-the-art works focused on proposing evasion strategies, only a few pointed out and tested possible countermeasures against such attacks. In the following, we summarize the proposed mitigation approaches with the same organization proposed in Section~\ref{sec:adversarial:subsec:evasion}. 

\myparagraph{Detection of Optimization-based Attacks} The only proposed approach to detect white- and gray-box gradient-based attacks against PDF files has been proposed by Smutz~\etal~\cite{smutz16-ndss}, who proposed an improvement of the common tree-based ensemble detection mechanism used for PDF detectors. By considering the voting results of each tree-based component of the ensemble, the authors defined a region of \emph{uncertainty} in which they classified evasion samples that obtained a score between specific thresholds. In this way, albeit the same samples were not explicitly regarded as malicious, the uncertain label would be enough to warn the user about possible evasion attempts. The attained results showed that \texttt{PDFRate-v2} performed significantly better than the previous \texttt{PDFRate-v1} at detecting the attack proposed by \Srndic and Laskov~\cite{srndic14-sp}.
    
\myparagraph{Detection of Heuristic-based Attacks} Concerning white-box mimicry, Smutz~\etal~\cite{smutz12-acsac} proposed a mitigation strategy in which the most discriminant features can be directly removed from the feature vector. However, such a choice can significantly impact the detection performances. To detect black-box reverse mimicry attacks, Maiorca~\etal~\cite{maiorca13-asiaccs,maiorca15-icissp,maiorca19-sp} proposed a combination of features that are extracted both from the structure and the content (in terms of embedded code) of the file, along with the extraction and separate analysis of embedded PDF files. In this way, it is possible to significantly mitigate such attacks (especially PDF Embedding ones, which are entirely detected), although more than $30\%$ of JS Embedding and EXE Embedding variants still managed to bypass detection~\cite{maiorca15-icissp,maiorca15-chapter,maiorca19-sp}. For the same problem, Smutz~\etal~\cite{smutz16-ndss} employed the same mitigation strategy proposed for detectin optimization-based attacks, and showed that the detection of reverse mimicry attacks (in particular, JS Embedding) significantly improved. An exception to such an improved detection is the PDF embedding variant, which still manages to evade the system, as even \texttt{PDFRate-v2} does not perform any analysis of possible embedded PDF payloads. 

Despite previous efforts at detecting evasion attacks, it is clear that a bulletproof solution to detect such attacks has not yet been developed. Accordingly, we propose three research guidelines that can be further expanded and applied both against optimization- and heuristic-based attacks. 

\myparagraph{Parsing Reliability} Proper parsing should always include, among the others: \emph{(i)} the extraction of the embedded content (that should be analyzed separately, either with the same or with different detectors); \emph{(ii)} using consolidated parsers (\eg, third-party libraries or, in general, parsers whose specifications are the closest to the official specs); \emph{(iii)} robustness against malformed files (including benign ones) that may make the parser crash. Further research could focus on implementing these three characteristics and discussing their impact on the overall detection performances.

\myparagraph{Feature Engineering} Following the results obtained by Maiorca~\etal~\cite{maiorca13-asiaccs,maiorca15-icissp,maiorca19-sp}, it is clear that a proper feature engineering would drastically increase the robustness of the learning systems against black- and white-box attacks. For this reason, the key research point to be followed should be designing features that are hard to be modified from the perspective of the attacker. Dynamic features are generally more robust, as the attacker should change the file behavior in order to change the feature value. However, when it is not possible or feasible to analyze PDF files dynamically, combining different types of static features can be a winning solution that may significantly increase the effort that is required by the attacker to evade the target systems.

\myparagraph{Robust and Explainable Learning Algorithms} Research should push on developing algorithms that are more robust against perturbations made by attackers. To this end, explicit knowledge of the attack model has to be considered during the design of the learning algorithm.
One possibility is to use robust optimization algorithms or, more generally, game-theoretical models. Such models have been widely studied in the literature of adversarial machine learning~\cite{globerson06-icml,bruckner12-jmlr}, largely before the introduction of \emph{adversarial training}~\cite{goodfellow15-iclr}, which essentially follows the same underlying idea.
Under this game-theoretical learning framework, the attacker and the classifier maximize different objective functions.\footnote{For the sake of completeness, it is worth pointing out that, in robust optimization, they maximize the same objective with opposite sign, yielding a zero-sum game} In particular, while the attacker manipulates data within certain bounds to evade detection, the classifier is iteratively retrained to correctly detect the manipulated samples~\cite{globerson06-icml,bruckner12-jmlr,bulo17-tnnls,biggio18-pr}. We argue that game-theoretical models may be helpful to improve robustness of PDF malware detectors.
Explainable machine-learning algorithms may provide another useful asset for system designers to analyze the trained models and understand their weaknesses. This approach has been recently adopted to inspect the security properties of learning systems and highlight their vulnerabilities~\cite{melis18-eusipco,demetrio19-itasec,guidotti19-csur,ribeiro16,baehrens10-jmlr}. In particular, it has been already observed (\eg, in spam filtering~\cite{kolcz09,biggio10-ijmlc} and Android malware detection~\cite{demontis17-tdsc}) that learning algorithms may tend to overemphasize some features, and that this facilitates evasion by a skilled attacker who carefully manipulates only those feature values. Developing specific learning approaches that distribute feature importance more evenly can significantly improve robustness to such adversarial evasion attempts~\cite{kolcz09,biggio10-ijmlc,demontis17-tdsc,melis18-eusipco}. This has been clearly demonstrated by Demontis~\etal~\cite{demontis17-tdsc} in the context of adversarial Android malware detection. In that work, the authors have shown that classifiers trained using a principled approach based on robust optimization eventually provide more-evenly distributed feature weights and much stronger robustness guarantees (even than the robust deep networks developed for the same task in~\cite{grosse17-esorics}).
Another defensive mechanism against evasion attacks consists of \emph{rejecting} anomalous samples at test time, similarly to the uncertainty-region approach defined by Smutz~\etal~\cite{smutz16-ndss}. More specifically, when a test sample is assigned an anomalous score in comparison to the scores typically assigned to benign or malware samples, it can be flagged as anomalous or classified as suspicious, requiring further manual inspection (see, \eg, the work in~\cite{melis17-vipar} for an example on adversarial robot vision).
Defensive mechanisms have also been proposed against poisoning and privacy attacks, respectively based on training data sanitization and robust learning, and on leveraging differentially-private mechanisms for learning private classifiers (we refer the reader to~\cite{biggio18-pr} for further details).

We finally argue that the aforementioned guidelines are not only useful for PDF malware detectors, but they can be extended to other malware detection systems that may be targeted by adversarial attacks. We believe that a correct application of the principles described above can be genuinely beneficial and can constitute a substantial aid to increase security of such systems.

\section{Conclusions}
\label{sec:conclusion}

In this paper, we presented a survey of malicious PDF detection in adversarial environments, featuring multi-folded contributions. First, we described how malicious PDF files perform their attacks in the wild. Accordingly, we described all machine learning-based solutions for PDF malware detection that have been proposed in the last decade. Notably, this part differentiates from previously proposed surveys, such as the ones by Nissim~\etal~\cite{nissim15-cose} and Elingiusti~\etal ~\cite{,elingiusti18-chapter}, which did not focus on those system components that are crucial to understanding adversarial attacks. For example, our work provided a deep insight into the pre-processing of PDF files, which has been exploited by many adversarial attacks. Second, we provided a comprehensive overview of the adversarial attacks that have been carried out against PDF malware detectors. In particular, we categorized the actual adversarial attacks against PDF detectors under a unifying framework and described them by also sketching possible novel strategies that can be employed by attackers. To this end, we employed a methodology similar to previous work on command-and-control botnets~\cite{gardiner16-csur}. However, PDF malware detection is clearly a different task and, to the best of our knowledge, this is the first work that thoroughly described adversarial machine learning applied to this field. The significant adversarial-related traits of this survey also completely distinguish it from other works on data mining for malware detection; for example, the work in \cite{ye17-csur} is not focused on the adversarial aspects of the problem. Despite the work by Cuan~\etal~\cite{cuan18-phd} pointed out some basics aspects of how adversarial machine learning has been applied to the detection of malicious PDF files, our work expands and discusses such problem in much wider details. Finally, we discussed existing mitigation strategies and sketched new research directions that could allow detecting not only current adversarial attacks but also novel ones.

Notably, the main goal of this work is that of discussing how PDF malware analysis has brought, across these recent years, significant advancements in understanding how adversarial machine learning could be applied to malware detection. The recent results described in this work have been beneficial in both research fields. On the one hand, research demonstrated that adversarial attacks constitute a concrete, real emerging security threat that can be extremely dangerous, as machine learning is now widely employed even by standard anti-malware solutions. On the other hand, discoveries concerning adversarial attacks pushed developers and security analysts to develop better protections and to explore novel information about malware that can be useful for classification. We believe that future work should focus on a more rigorous application of security-by-design principles when building detection systems. In this way, it will be possible to offer real, active mitigation of the numerous evasion attacks that are being progressively used in the wild.

\section*{Acknowledgments}
This work was partially supported by the INCLOSEC (funded by Sardegna Ricerche - CUP G88C1700
0080006) and PISDAS (funded by Regione Autonoma della Sardegna -  CUP E27H14003150007) projects; and by the EU H2020 project ALOHA, under the European Union's Horizon 2020 research and innovation programme (grant no. 780788).

\bibliographystyle{ACM-Reference-Format}
\bibliography{pdf-biblio,misc,bibDB}

\end{document}